\documentclass[prb,aps,showpacs]{revtex4}

\usepackage{graphicx}
\usepackage{ifthen}
\usepackage{ifpdf}
\usepackage{color}
\definecolor{red}{rgb}{1.,0.,0.}

\usepackage{latexsym}
\usepackage{amssymb}
\usepackage{bm}
\newcommand{\half}{\mbox{\small $\frac{1}{2}$}}

\newcommand{\abs}[1]{\left|#1\right|}
\newcommand{\im}{\mbox{Im}}
\newcommand{\re}{\mbox{Re}}
\newcommand{\eexp}{\mbox{e}^}
\newcommand{\bra}{\left\langle}
\newcommand{\ket}{\right\rangle}
\newcommand{\Exp}[1]{\mbox{Exp} {\left[ #1 \right]}}

\newcommand{\beq}[1]{\begin{eqnarray}\ifthenelse{#1=-1}{\nonumber}
{\ifthenelse{#1=0}{}{\label{e#1}}}}
\newcommand{\eeq}{\end{eqnarray}}
\newcommand{\be}[1]{\begin{eqnarray}}
\newcommand{\ee}{\end{eqnarray}}

\newcommand{\hide}[1]{}
\newcommand{\D}{\mathcal{D}}
\newcommand{\avg}[1]{\left\langle #1 \right\rangle}
\newcommand{\order}[1]{\mathcal{O} {\left( #1 \right)}}
\newcommand{\sign}{\mbox{sign}}

\begin{document}

\title{Rings and Coulomb boxes in dissipative environments}
\author{Yoav Etzioni{$^1$}, Baruch Horovitz{$^1$} and Pierre Le Doussal{$^2$} }

{\affiliation{{$^1$} Department of Physics, Ben Gurion University,
Beer Sheva 84105 Israel}
 \affiliation{{$^2$} CNRS-Laboratoire de
Physique Th{\'e}orique de l'Ecole Normale Sup{\'e}rieure, 24 rue
Lhomond,75231 Cedex 05, Paris France.}

\begin{abstract}
We study a particle on a ring in presence of a dissipative Caldeira-Leggett environment and derive its response to a DC field. We show how this non-equilibrium response is related to a flux averaged equilibrium response.
We find, through a 2-loop renormalization group analysis, that a large dissipation parameter $\eta$ flows to a fixed point $\eta^R=\hbar/2\pi$. We also reexamine the mapping of this problem to that of the Coulomb box and show that the relaxation resistance, of recent interest, is quantized for large $\eta$. For finite $\eta>\eta^R$ we find that a certain average of the relaxation resistance is quantized. We propose a Coulomb box experiment to measure a quantized noise.

\end{abstract}


\maketitle

\section{Introduction}

Two of the most important mesoscopic structures are rings, for the study of persistent currents, and quantum dots or boxes, for the study of charge quantization. Of particular recent interest is the quantization of the relaxation resistance, defined via an AC capacitance of a single electron box (SEB). An SEB is defined as a quantum dot that has $N_c$ transmission channels into a single electron reservoir, i.e. an electrode, and is capacitively coupled to a gate voltage. This setup is equivalent to an RC circuit \cite{buttiker,gabelli} whose capacitance at low frequency $\omega$ has the form $C_0(1+i\omega C_0R_q)$, identifying the relaxation resistance $R_q$.
Following the prediction of B\"{u}ttiker, Thomas and Pr\^{e}tre \cite{buttiker} that $R_q=h/2e^2$ for a single channel, a quantum mesoscopic RC circuit has been implemented in a two-dimensional electron gas \cite{gabelli} and $R_q=h/2e^2$ has been measured. The theory has been recently extended to include Coulomb blockade effects \cite{mora,filippone} showing that $R_q=h/2e^2$ is valid for small dots and crosses over to $R_q=h/e^2$ for large dots.

In parallel, recent data has observed Aharonov-Bohm oscillations from single electron states in semiconducting rings \cite{kleemans}. Further theoretical works have considered the effects of dissipative environments on a single particle in a ring \cite{guinea}, in particular studying the renormalization of the mass $M^*$ and its possible relation to dephasing \cite{guinea,golubev1,kagalovsky,bh}. A related case of a ring coupled by tunneling to an electron lead has also been studied \cite{arrachea}.

It is rather remarkable that the ring and box problems are related via the Ambegaokar, Eckern, and Sch\"{o}n (AES) mapping \cite{AES} where the ring experiences a Caldeira-Leggett (CL) \cite{CL} environment. While the exact mapping assumes weak tunneling into the box with many channels, it has been extensively used to describe various tunnel junctions \cite{schon}, the Coulomb blockade phenomena in SEB and in the single electron transistor (SET) \cite{schon,golub,falci,hofstetter,herrero,bulgadayev,lukyanov1,lukyanov2,burmistrov1,burmistrov2}.

\begin{figure}[b]
\begin{center}
\includegraphics[scale=0.4]{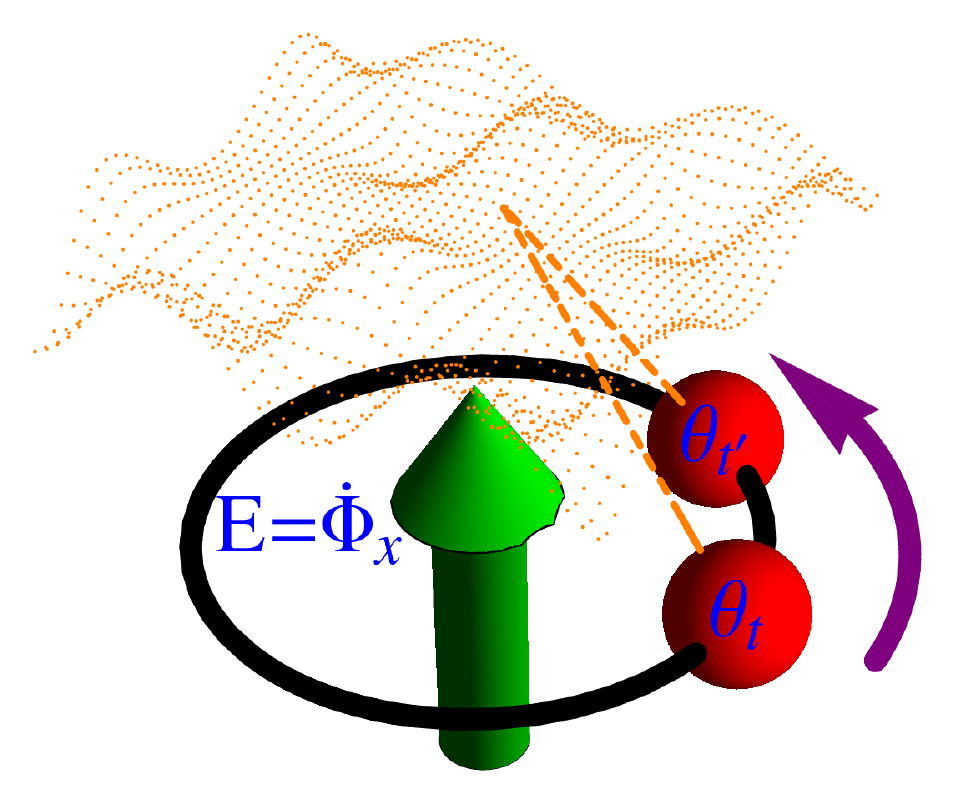}
\end{center}
\caption{Artist view of a particle on a ring, coupled to an environment, with a field $E=\dot{\phi}_x$ due to a time dependent flux through the ring. The particle polarizes the environment which in turn modifies the motion of the particle at later times, i.e. an effective non-local interaction.}
\end{figure}

The ring problem is defined by a particle confined to a ring, coupled to a dissipative environment of the Caldeira-Leggett type, and in presence of a field $E$, generated by a time dependent flux $\phi_x$ through the ring. This scenario is schematically illustrated in Fig.1. The Caldeira-Leggett coupling can be realized, e.g., by a normal metal whose mean free path is much larger than the ring's radius \cite{bh}.
In the present work we address the ring problem by the real time Keldysh method and study it using a 2-loop expansion and renormalization group (RG) reasoning. We find that perturbation theory identifies an unexpected new small parameter $\sin(\frac{\hbar}{2\eta})$ where $\eta$ is the dissipation parameter on the ring, or the lead-dot coupling in the SEB. We infer that a large $\eta$ flows to a fixed point $\eta^R$ with $\hbar/2\eta^R=\pi$.  While the thermodynamics of the ring type problem has been much studied, including extensive Monte Carlo studies \cite{herrero,lukyanov2} of $M^*$, no sign of a finite coupling fixed point has been detected. Our method evaluates the response to a strictly DC electric field $E$, equivalent to a magnetic flux through the ring that increases linearly with time, hence a non-equilibrium response. We claim that thermodynamic quantities like $M^*$, that are flux sensitive, decouple from the response to $E$, a response that averages over flux values. This general relation between non-equilibrium and equilibrium responses is given by Eq. (\ref{e39}) below. This relation has been noticed for a model with particle tunneling between a ring and an environment \cite{buttiker2}.

In terms of the SEB, our results extend the previous analysis \cite{mora,filippone} to the case of many channels $N_c$, an experimentally realizable scenario \cite{devoret}. We note that for $N_c > 1$ the relaxation resistance for noninteracting electrons \cite{buttiker} becomes $h/(2N_ce^2)$. We find that for strong coupling, $\eta/\hbar \gtrsim 1$ the relaxation resistance is quantized to $e^2/h$ up to an exponentially small correction $\sim \eexp{-\pi\eta/\hbar}$. For finite $\eta$, but still $\eta>\eta^R$ we find that a certain average of the relaxation resistance is quantized (see Eq. \ref{e82}).

The present work considerably expands our previous letter \cite{etzioni}.
In section II we present the ring and box models, with some exact general properties. In section III we present RG and numerical solutions for the semiclassical case, while section IV presents the perturbation and RG analysis of the full quantum case. The discussion section V summarizes our results, discusses its topological interpretation and details a proposed Coulomb box experiment to detect our predicted quantized noise. The Appendices give details of the ring-box mapping and of the various perturbation expansions. We consider temperature $T=0$ throughout.

 As a simple motivation for our main result, we present here a topological interpretation of the fixed point $\eta^R$, based on the Thouless charge pump concept \cite{thouless}. Consider a slow change of $\phi_x$ by one unit with $\hbar{\dot \phi}_x=\eta^R\langle{\dot\theta}\rangle$. For the special value $\eta^R=\hbar/(2 \pi)$ the total change in the position of the particle is
$\int_t \langle{\dot\theta}\rangle dt=2\pi$, i.e. the particle comes back to the same position on the ring and a unit charge has been transported.

\section{The model and general properties}

\subsection{Semiclassical model}
We derive first a Langevin equation for a particle on a ring.
Consider the standard Langevin equation for a particle with coordinate $x_t$ in one-dimension of the form
\beq{01}
R^{-1}_{t,t'}x_{t'}=\xi_t
\eeq
where $\xi_t$ is a Gaussian random force from an environment, where the average on the environment degrees of freedom is
\beq{02}
\langle\xi_t\xi_{t'}\rangle=B_{t,t'}
\eeq
This relation defines a linear response for either $x_\omega=R_\omega\xi_\omega$ or $\xi(\omega)=R^{-1}(\omega)x(\omega)$, after Fourier transforms, e.g. $R(\omega)$ is the Fourier transform of $R_t=R_{t,0}$. Hence the fluctuation dissipation theorem (FDT) at temperature $T$ can be applied either way, leading to
\beq{03} K_x(\omega)&=&\hbar\coth
(\half\beta\hbar\omega)\im[R_\omega]\nonumber\\
B_\omega&=&\hbar\coth
(\half\beta\hbar\omega)\im \frac{-1}{R_\omega}
\eeq
where $K_x(\omega)$ is the Fourier transform of $K_x(\tau)=\half\langle x_tx_{t+\tau}+
 x_{t+\tau}x_t\rangle$. The simplest choice corresponds to a particle with mass $m$ and a friction coefficient $\eta$, so that at temperature $T=0$
 \beq{04}
 m\ddot{x}_t+\eta \dot{x}_t&=&\xi_t\nonumber\\
 R_0(\omega)&=&\frac{-1}{m\omega^2+i\omega\eta}\qquad   R_0(t)=\frac{1}{\eta}[1-\eexp{-\eta
t/m}]\Theta(t) \nonumber\\
 B_\omega&=&\hbar\eta |\omega|\qquad  B_t=\frac{-\hbar\eta}{\pi t^2}\,\,(t\neq 0)
  \eeq
where $R_0(t-t')$ is the response in this case. While the mass provides a high frequency cutoff which we denote $\omega_c=\eta/m$, the singularity of $B(t)$ at $t=0$ implies the need for an additional cutoff. This additional cutoff is a convenience and will be used below in the simulations as well as in the RG derivation. A method for deriving general response functions is based on Kramers Kronig relations \cite{ford}. In the notation of Eq. (2.7) of Ref. [\onlinecite{ford}] we choose $\re \mu(\omega)=\eta/(1+\omega^2\tau_0^2)$ so that the response function $R^{-1}_{t-t'}$, after Fourier, is
\beq{05}
R^{-1}_\omega=-m\omega^2-\frac{i\omega\eta}{1-i\omega\tau_0}\,. \eeq
To justify the use of this form it suffices to say that it
 has the remarkable and necessary property
that both $R_\omega$ and $R^{-1}_\omega$ have no poles in the
upper half plane, as needed for causal functions; [note that $R_\omega$ reduces to $R_0(\omega)$ when $\tau_0=0$]. The FDT at $T=0$ gives
\beq{06}
B_\omega=\frac{\hbar|\omega|\eta}{1+\omega^2\tau_0^2}\eeq
so that $1/\tau_0$ provides a cutoff on the environment frequencies, in addition to the cutoff $\frac{m}{\eta}=\frac{1}{\omega_c}$.
Hence for $4\tau_0<m/\eta$, ($\delta\rightarrow +0$)
\beq{07}
R_t&=&\Theta(t)\frac{\tau_0}{m}[\frac{m}{\eta\tau_0}\eexp{-\delta t}+\frac{1-\lambda_1}{\lambda_1 x}
\eexp{-\lambda_1t/\tau_0}-\frac{1-\lambda_2}{\lambda_2 x}
\eexp{-\lambda_2t/\tau_0}]\nonumber\\
\lambda_1&=&\half[1+x], \qquad \lambda_2=\half[1-x], \qquad x=\sqrt{1-\frac{4\eta\tau_0}{m}} \eeq
while for $4\tau_0>m/\eta$ with $x=\sqrt{\frac{4\eta\tau_0}{m}-1}$
\beq{08}
R_t=\Theta(t)\frac{1}{\eta}\{\eexp{-\delta t}-[\frac{1-x^2}{2x}\sin (xt/2\tau_0)+\cos (xt/2\tau_0)]
\eexp{-t/2\tau_0}\}
\eeq

Consider now the two-dimensional system and its projection on a ring, i.e. ${\bf x}_t=(\cos\theta_t,\sin\theta_t)$ so that $\theta_t$ is the angular position of the particle and the radius is chosen as unity. In cartesian coordinates we define random forces in the $x,y$ directions so that  $R^{-1}_{t-t'}\cos\theta_{t'}=-\xi^b_t,\,
R^{-1}_{t-t'}\sin\theta_{t'}=\xi^a_t$. The ring potential confines the motion to
the azimuthal part, so that only the tangent force
$-\xi^a_t\cos\theta_t+\xi^b_t\sin\theta_t$ is allowed, hence
\beq{09}
-\sin\theta(t)R^{-1}_{t-t'}\cos\theta_{t'} +
\cos\theta_t R^{-1}_{t-t'}\sin\theta_{t'}
=\xi^a_t\cos\theta_t+\xi^b_t\sin\theta_t +E
\eeq
where $\xi^a_t,\,\xi^b_t$ are independent and each having the correlations of Eq. (\ref{e02}).
An external tangent electric field $E$ has been added corresponding to a flux through the ring that is increasing linearly with time $\phi_x=Et$. With
$R_0(t-t')$ given by Eq. (\ref{e04}) the differential form
$R^{-1}_0(t)=mr\partial_t^2+\eta r\partial_t$, can be used leading to
\beq{10}
m\ddot{\theta}_t +\eta \dot{\theta}_t=\xi^a_t
\cos\theta_t +\xi^b_t \sin\theta_t  +E
\eeq
This nonlinear Langevin equation has been studied also in the SET context \cite{golubev2}.
Comparing the time derivatives in Eq. (\ref{e10}) identifies a cutoff frequency $\omega_c=\eta/m$. At $\omega>\omega_c$ the mass term dominates while at $\omega<\omega_c$ the environment dominates, leading to renormalizations.
The nonlinear Langevin's equation (\ref{e10}), including an average on the random forces, is equivalent to a partition function
\beq{11}
&\ & Z = \int \D[\theta,\xi] \
\delta \left( m \ddot{\theta}_t + \eta \dot{\theta}_t -
\xi^a_t \cos \theta_t - \xi^b_t \sin \theta_t - E \right) \eexp{-\int_{\omega}[\abs{\xi^a_\omega}^2+\abs{\xi^b_\omega}^2]/2 B_\omega}\nonumber\\
\eeq
Introducing the 'quantum' field $\hat{\theta}$ by
$\delta(X_t) = \int \D [\hat{\theta}]
\eexp{i \hat{\theta}_t X_t}$, and averaging over the noise field $\xi_x,\xi_y$ results in
the semi classical partition function
$Z = \int \D[\theta,\hat\theta] \eexp{-S[\theta,\hat\theta]}$
where $S[\theta,\hat\theta] = S_0 + S_{int}$
is given by the $t,t'$ integrations
\beq{12}
&\ & S_0 = i \int_{t,t'} \hat{\theta}_t (R_{t,t'})^{-1} \theta_{t'} -
i E \int_{t'} \hat{\theta}_{t'}  =
i \int_\omega R_\omega^{-1} \hat{\theta}_\omega {\theta}_{-\omega}   -
i E \int_{t'} \hat{\theta}_{t'} \nonumber\\
&\ & S_{int} = \frac12\int_{t,t'} \hat{\theta}_t B_{t,t'} \hat{\theta}_t \cos(\theta_t-\theta_{t'}).
\eeq
This has the form of a Keldysh action, with $\theta,\hat\theta$ being the classical and quantum fields, respectively. We will see below that this action is the semiclassical $\hbar\rightarrow 0$ limit of the full quantum system.

\subsection{Quantum model}

We proceed to define the full quantum problem. The one-dimensional Langevin system \cite{CL,hakim,fisher}
 has the Keldysh partition $Z=\int  {\cal D}{\hat x}_t{\cal D}x_t\eexp{-S_K}$ where
\beq{13} S_K=i\int_{t,t'} {\hat x}_t R^{-1}_{t,t'}x_{t'} +\half
\int_{t,t'}{\hat x}_tB_{t,t'}{\hat x}_{t'} \eeq
and ${\hat x}_t,\,x_t$ are the quantum and classical fields, respectively,
\beq{14} x_t=\half(x^+_t+x^-_t)\qquad
{\hat x}_t=\frac{1}{\hbar}(x^+_t-x^-_t)\eeq
and $x_t^{\pm}$ are on the upper and lower Keldysh contour, respectively. On a ring, we use a 2-dimensional vector notation
\beq{15} {\bf x}_t^+=[\cos\theta^+_t,\sin\theta^+_t]\qquad
{\bf x}_t^-=[\cos\theta^-_t,\sin\theta^-_t]\eeq
Defining
\beq{16} \theta_t=\half(\theta^+_t+\theta^-_t)\qquad
{\hat \theta}_t=\frac{1}{\hbar}(\theta^+_t-\theta^-_t)\eeq
 and using trigonometric identities we obtain the quantum action
\beq{17}S_K=i\frac{2}{\hbar}\int_{t,t'}R^{-1}_{t,t'}\sin(\frac{\hbar}{2}{\hat
\theta_t})\cos(\frac{\hbar}{2}{\hat
\theta}_{t'})\sin(\theta_{t'}-\theta_t)+\frac{2}{\hbar^2}\int_{t,t'}B_{t,t'}\sin(\frac{\hbar}{2}{\hat
\theta_t})\sin(\frac{\hbar}{2}{\hat
\theta}_{t'})\cos(\theta_{t'}-\theta_t)\nonumber\\ \eeq

We note that the path integral involves continuous $\theta_t$ trajectories that can involve $n$ rotations around the ring. Consider the time evolution from an initial wavefunction $\psi(\theta_0,t_0)$ at time $t_0$ to a final state
$\psi({\tilde\theta}_t,t)$, where both initial and final angles are compact, $0<\theta_0,{\tilde\theta}_t<2\pi$,
\beq{18}
\psi({\tilde\theta}_t,t)=\int_0^{2\pi}d\theta_0\sum_n\int_{\theta_0}^{{\tilde\theta}_t+2\pi n}{\cal D}\theta
\eexp{-S(t,t_0)}\psi(\theta_0,t_0)
\eeq
The sum on the integers $n$ expresses that the probability to arrive at a given ${\tilde\theta}_t$ is a sum
of probabilities, each with n rotations. The path integral can therefore be written in terms of a decompactified variable $\theta_t={\tilde\theta}_t+2\pi n$, i.e. $\sum_n\int_{\theta_0}^{{\tilde\theta}_t+2\pi n}{\cal D}\theta
\rightarrow \int_{\theta_0}^{\theta_t}{\cal D}\theta$ where now $-\infty<\theta_t<\infty$. This shift does not affect the periodic forms in (\ref{e17}), however it does affect an external electric field $E$. Consider a time dependent flux $\phi_x(t)=Et$ that contributes to the action a term $\int_{t_i}^{t_f}\phi_x(t){\dot\theta}_tdt=
-E\int_{t_i}^{t_f}\theta_tdt+\phi_x(t_i)\theta_{t_i}-\phi_x(t_f)\theta_{t_f}$. The partial integration is allowed only for the decompactified variable $\theta_t$, i.e. the work done by $E$ is finite for each $2\pi$ rotation. The boundary terms are neglected, e.g. one can choose $\phi_x(t_i)=\phi_x(t_f)=0$ where $t_i,t_f\rightarrow -\infty$ are boundary times on a Keldysh contour; the field $E$ is turned on slowly away from these times.

In the following we will consider a perturbative scheme with a field $E$ and a bare velocity
$v=E/\eta$ and $\theta_t$ is decomposed to $\theta_t=\delta\theta_t+vt$; (the true velocity is defined below as $v^R(E)=\langle\dot\theta_t\rangle$). The velocity $v$ provides a low frequency cutoff eliminating divergence of the perturbative expansion and eventually allows for RG treatment.
It will be convenient to use the two-cutoff response Eq. (\ref{e05}) with $R^{-1}_\omega=-m\omega^2+\delta R^{-1}_\omega$, where $\delta R^{-1}_\omega=\frac{-i\omega \eta}{1-i\omega\tau_0}$, hence
\beq{19}
\delta R^{-1}_{t,t'}=\partial_{t'}\int_{\omega}\frac{-\eta}{1-i\omega\tau_0}\eexp{-i\omega(t-t')}=
-\frac{\eta}{\tau_0}\partial_{t'}[\eexp{-(t-t')/\tau_0}\Theta(t-t')]=
\frac{\eta}{\tau_0}\eexp{-(t-t')/\tau_0}\Theta(t-t')\partial_{t'}
\eeq
The operator identity is satisfied for any function decaying faster then $\eexp{|t'|/\tau_0}$ at $t'\rightarrow -\infty$. Note,
\beq{20}
i\int_{t,t'}{\hat\theta}_t\delta R^{-1}_{t,t'} vt'=i\frac{\eta v}{\tau_0}\int_{t}{\hat\theta}_t
\int_{-\infty}^t \eexp{-(t-t')/\tau_0}dt'=iv\eta\int_t {\hat\theta}_t
\eeq
The mass term with $m\omega^2\rightarrow \delta(t-t')\partial_t\partial_{t'}$ produces $m\int_t \dot{\hat
\theta}_t{\dot \theta}_t=m\int_t\dot{\hat\theta}_t\delta{\dot \theta}_t+mv\int_t\dot{\hat\theta}_t$; the last term with $mv=E/\omega_c$ is neglected relative to the field term $\int_t Et\dot{\hat\theta}_t$. The full action is then
\beq{21}
S_K&=&S_0+S_{int}+S_c\nonumber\\
S_0&=& i\int_{t,t'} {\hat\theta}_t R^{-1}_{tt'} \theta_{t'} - i E \int_t {\hat\theta}_t =i\int_{t,t'} {\hat\theta}_t R^{-1}_{tt'}\delta\theta_{t'}\nonumber\\
S_{int}&=&\frac{2}{\hbar^2}\int_{t,t'}B_{t,t'}\sin(\frac{\hbar}{2}{\hat
\theta_t})\sin(\frac{\hbar}{2}{\hat \theta}_{t'})\cos(\theta_{t'}-\theta_t)\nonumber\\
S_c&=&i\frac{2}{\hbar}\int_{t,t'}\delta R^{-1}_{t,t'}[\sin(\frac{\hbar}{2}{\hat
\theta_t})\cos(\frac{\hbar}{2}{\hat\theta}_{t'})\sin(\theta_{t'}-\theta_t)-
\frac{\hbar}{2}{\hat\theta}_t\theta_{t'}]
\eeq

The use of a single cutoff  Eq. (\ref{e04}) with
\beq{22}
R_0^{-1}(t,t')=\delta(t-t')[m\partial_t\partial_{t'}+\eta\partial_{t'}]
\eeq
leads to a simpler action.  It corresponds to $\tau_0\rightarrow 0$, hence  $\delta R^{-1}_{t,t'}\rightarrow \eta\delta(t-t')\partial_{t'}$,
\beq{23} \frac{2}{\hbar}R_0^{-1}(t,t')\sin(\frac{\hbar}{2}{\hat
\theta}_{t})\cos(\frac{\hbar}{2}{\hat
\theta}_{t'})\sin(\theta_{t'}-\theta_t)=\delta(t-t')[m{\dot{\hat
\theta}}_t{\dot \theta}_t+\frac{\eta}{\hbar}\sin(\hbar{\hat\theta}_t){\dot \theta}_{t^-}]
\eeq
where $t^-$ is infinitesimal below $t$ so that the retarded nature of $R^{-1}_{t,t'}$ is maintained.
The action $S_K=S_0+S_{int}+S_c$ is then
\beq{24}
S_0&=&i\int_{t,t'} {\hat\theta}_t R^{-1}_0(t,t')\delta\theta_{t'} = i\int_t
[m{\dot{\hat\theta}}_t\delta{\dot \theta}_t+\eta{\hat\theta}_t\delta{\dot \theta}_t]=
i\int_t[m{\dot{\hat\theta}}_t{\dot \theta}_t+\eta{\hat\theta}_t{\dot \theta}_t]- i E \int_t {\hat\theta}_t
 \nonumber\\
S_{int}&=&\frac{2}{\hbar^2}\int_{t,t'}B_{t,t'}\sin(\frac{\hbar}{2}{\hat
\theta_t})\sin(\frac{\hbar}{2}{\hat \theta}_{t'})\cos(\theta_{t'}-\theta_t)\nonumber\\
S_c&=&\frac{i\eta}{\hbar}\int_t[\sin(\hbar{\hat\theta}_t){\dot\theta}_{t^-}
-\hbar{\hat\theta}_t{\dot\theta}_{t^-}] \qquad\qquad \tau_0\rightarrow 0.
\eeq
Note that this action reduces to that to the semiclassical case Eq. (\ref{e12}) when $\hbar\rightarrow 0$.

\subsection{Renormalized friction}

The renormalized friction $\eta^R(E)$ is defined by the renormalized response $R^R_{t,t'}=i\bra\theta_t{\hat\theta}_{t'}\ket_E$ and its DC limit:
\beq{25}
\frac{1}{\eta^R(E)}=\lim_{\omega\rightarrow 0}(-i\omega R^R_\omega)
\eeq
in analogy with the bare form Eq. (\ref{e04}). We show now that the renormalized $\eta^R(E)$ is also the local slope of $\frac{dv^R}{dE}$, where $v^R$ is the $E$ dependent renormalized velocity
\beq{26}
v^R\equiv \bra{\dot \theta}_t\ket=\int\D[\theta] {\dot \theta}_t\eexp{-S_K}
\eeq
Therefore
\beq{27}
\frac{dv^R}{dE}&=&i\bra \int_{t'}{\dot\theta}_t{\hat\theta}_{t'}\ket=\int_{t'}\frac{d}{dt}R^R_{t,t'}=
\int_{t'}\int_{\omega}(-i\omega)R^R_\omega \eexp{-i\omega(t-t')}=\lim_{\omega\rightarrow 0}\frac{-i\omega}
{-i\eta^R(E)\omega}=\frac{1}{\eta^R(E)}\nonumber\\
\eeq
In particular we are interested in the limit $\eta^R=\eta^R(E\rightarrow 0)$.

We show now an alternative procedure for evaluating $\eta^R$. Consider the Keldysh partition $Z=\int\D[\theta]\eexp{-S_K}$ and shift ${\hat \theta}_t\rightarrow {\hat \theta}_t+a_t$. The result must be $a_t$ independent, and choosing the form (\ref{e23}) with $\tau_0\rightarrow 0$ (the following identity is actually independent of cutoff choices)
\beq{28}
0&=&\frac{\delta Z}{\delta a_t}|_0=-\langle \frac{\delta (S_0+S_{int}+S_c)}{\delta{\hat\theta}_t}\rangle=
-i(\eta v^R-E-\delta E)\nonumber\\
&&\delta E\equiv i\bra\frac{\delta (S_{int}+S_c)}{\delta{\hat\theta}_t}\ket
\eeq
since $-i\langle \frac{\delta S_0}{\delta{\hat\theta}_t}\rangle=-m\langle \ddot\theta_t\rangle+\eta\langle\dot\theta\rangle -E$ and $v^R$ is time independent, at least for long times.

Taking an $E$ derivative of Eq. (\ref{e28}) and using (\ref{e27}) we obtain
\beq{29}
\frac{1}{\eta^R(E)}=\frac{1}{\eta}+\frac{1}{\eta^2}\frac{\partial}{\partial v}\delta E
\eeq
We have checked, up to 2nd order terms, that the results of (\ref{e27}) and (\ref{e29}) coincide. The use of (\ref{e29}) is technically easier.

\subsection{Equilibrium correlations}
In this section we consider the equilibrium response to a change in flux and derive a relation with the nonequilibrium response to a field.

 Consider now the form of ${\tilde K}(\omega)$ as a response to a flux $\phi_x$. Linear response to
 $\delta{\cal H}_{ring}=+\hbar{\dot \theta}\delta\phi_x(t)$ is
 \beq{30}
 \hbar\langle {\dot\theta}\rangle=-\int_{t'}{\tilde K}_{t,t'}\delta\phi_x(t')
 \eeq
 This corresponds also to the velocity correlation
 \beq{31}
 {\tilde K}_{t,t'}&=&+i\theta(t-t')\langle [{\dot \theta}_t,{\dot\theta}_{t'}]\rangle
 \eeq

 We expect that the DC response is positive for small $\phi_x$, hence define
 \beq{32}
 {\tilde K}(\omega)=-K_0(\phi_x)+i\omega K_1(\phi_x) +O(\omega^2)
 \eeq
 The response $K_0(\phi_x)$ is the persistent current, i.e. for a static flux one can integrate (\ref{e30})
 \beq{33}
 \langle {\dot\theta}\rangle=\int_0^{\phi_x}K_0(\phi_x')d\phi_x'
 \eeq
 The periodicity of the persistent current implies $\int_0^{1}K_0(\phi_x)d\phi_x=0$. The curvature of the free energy $F$ (or energy at $T=0$) at $\phi_x=0$ is a well studied object \cite{guinea,golubev1,kagalovsky,bh}. For general $\phi_x$ it is defined by a Matsubara imaginary time correlation
 \beq{34}
 \frac{1}{\hbar}\frac{\partial^2 F}{\partial \phi_x^2}=(\beta)^{-1}\int_0^{\beta}\int_0^{\beta}\langle {\dot\theta}_{\tau}{\dot\theta}_{\tau'}\rangle^c d\tau d\tau'= K_0(\phi_x)
 \eeq
 where $K(i\omega_n=0)=+K_0$ (there is a sign difference in the standard Matsubara notation). An effective mass is defined by $K_0(0)=\hbar/M^*$ so that $M^*=m$ without interactions, while for strong $\eta\gg 1$ coupling $M^*\sim \eexp{\pi\eta}$ is exponentially large \cite{guinea,golubev1,kagalovsky,bh}.

 To appreciate the role of $K_1$ consider FDT for the symmetrized correlation at small $\omega$
 \beq{35}
 \langle|{\dot\theta}_{\omega}|^2\rangle^{sym}={\mbox{\text sign}}
 \omega\cdot\im {\tilde K}_\omega=|\omega|K_1
 \eeq
 The diffusion involves the response $\langle |\theta_{\omega}|^2\rangle=K_1/|\omega|$, hence for $t\rightarrow \infty$
 \beq{36}
 \langle (\theta_t-\theta_0)^2\rangle=K_1\int d\omega\frac{1-\cos\omega t}{\pi|\omega|}=
 \frac{2K_1}{\pi}\ln (\omega_x t)
 \eeq
 where $\omega_x$  is a characteristic frequency where higher order terms in $\omega$ terms set in.

 Consider now the linear response to an electric field $\delta{\cal H}_{ring}=- E(t)\theta_t$ and use  the response  $\langle\theta_t\rangle=R^R_{t,t'}E(t')$ The definition (\ref{e25}) implies that the  low $\omega$ limit has the form $R^R_\omega=\frac{-1}{i\omega\eta^R}$.
 Since $E=\hbar{\dot\phi}_x$ we expect $\hbar\omega^2R^R_\omega={\tilde K}(\omega)$. However, there is a difficulty with the latter relation, if taken literally,
 \beq{37}
 \frac{-\hbar\omega^2}{i\omega\eta^R}?=?-K_0(\phi_x)+i\omega K_1(\phi_x)
 \eeq
 It is also not clear which $\phi_x$ to use in this relation.
 To resolve this issue consider the ${\tilde K}$ response with a constant electric field
 \beq{38}
 \hbar\langle{\dot\theta}_t\rangle= - \int_{t'}{\tilde K}_{t,t'}\cdot Et'
 \eeq
 Note first that an additional constant $\phi_x$ in $\frac{1}{\hbar}Et'+\phi_x$ can be eliminated by redefining the origin of the time $t'$, hence the persistent current part should be eliminated. More precisely, define $\phi_x(t)=\frac{1}{\hbar}Et$; the $\omega=0$ component $K_0(\phi_x)=K_0(\frac{1}{\hbar}Et)$ becomes a periodic function, i.e. an AC response with frequency $\omega_E=
 \frac{2\pi}{\hbar} E$. For $\omega\rightarrow 0$ this persistent current response averages to zero, i.e. $\int_0^1K_0(\phi_x)d\phi_x=0$. The same reasoning applies to a $\phi_x$ average on $K_1(\phi_x)$.
 Hence for the purpose of evaluating the DC response of (\ref{e25}) we need to average on the flux in (\ref{e32}), hence
 \beq{39}
 \lim_{E \to 0} \lim_{\omega \to 0}\frac{{\tilde K}(\omega)}{i\omega}=\int_0^1K_1(\phi_x)d\phi_x=\frac{\hbar}{\eta^R}\,.
 \eeq
 The order of limits in (\ref{e05}) signifies that $\eta^R$ is essentially a non-equilibrium response.  The equilibrium - nonequilibrium relation (\ref{e39}) has been noticed in solution of a Boltzmann relaxation equation for particles on a ring, allowing for particle tunneling into an environment \cite{buttiker2}.

 The physical picture is that in a DC field the particle rotates around the ring and produces two types of currents. First is the persistent current that oscillates in time as $\phi_x$ increases and is therefore time averaged to zero; this current is non-dissipative. Second, there is a genuine DC response from the $i\omega K_1$ term, which is dissipative.

 \subsection{The Coulomb box}
 Consider now the Coulomb box system, i.e. a finite region (a "dot") with charging energy $E_c$ coupled by tunneling to a single metallic lead.  The Hamiltonian is
 \beq{40}
 {\cal H}=\sum_k \epsilon_k a^{\dagger}_{k, i}a_{k, i}+\sum_{\alpha,i}\epsilon_{\alpha}d^{\dagger}_{\alpha, i}d_{\alpha, i}+E_c({\hat N}-N_0)^2 +\sum_{k,\alpha,i}t_{k,\alpha,i}a^{\dagger}_{k,i}d_{\alpha,i}+h.c.
 \eeq
 where  $i=1,...,N_c$ are channel indices, $d_{\alpha,i}$ are dot electron operators with spectra $\epsilon_{\alpha}$, $a_{k,i}$ are lead electron operators with spectra $\epsilon_k$,   ${\hat N}=\sum_{\alpha,i}d^{\dagger}_{\alpha,i}d_{\alpha,i}$ is the number operator on the dot, $E_c=e^2/2C_g$ is the charging energy with $C_g$ is the geometric (bare) capacitance, $N_0$ is the gate voltage in units of $2E_c$. The channel index $i$ is diagonal in the tunneling term, i.e. corresponds to transverse modes that are conserved in tunneling.

 Consider the density correlations
 \beq{41}
  \nonumber\\  K_{t,t'}&=&+i\theta(t-t')\langle [{\hat N}_t,{\hat N}_{t'}]\rangle
\eeq
The AES mapping to the ring problem is reproduced in Appendix A. In particular, $N_0$ corresponds to $-\phi_x$, $2E_c$ to $\hbar^2/m$ and the relation to the velocity correlation on the ring is
\beq{42}
\hbar^2{\tilde K}_{t,t'}&=&-2E_c\hbar\delta(t-t')+4E_c^2K_{t,t'}
\eeq
Using the notation \cite{mora} $K(\omega)=\hbar\frac{C_0}{e^2}(1+i\omega C_0R_q)$, where $C_0$ is the renormalized capacitance and $R_q$ is the relaxation resistance, we obtain
 \beq{43}
 \hbar{\tilde K}(\omega)=-2E_c+4E_c^2\frac{C_0}{e^2}(1+i\omega C_0 R_q)
 \eeq
 Hence the mapping between the Coulomb box and the ring for the curvature is, using (\ref{e34})
 \beq{44}
 \frac{\hbar^2}{M^*(\phi_x)}=\hbar K_0(\phi_x)=2E_c(1-\frac{C_0}{C_g})\qquad \Rightarrow
 \frac{m}{M^*(\phi_x)}=1-\frac{C_0(N_0)}{C_g}
 \eeq
 while for the dissipation, using (\ref{e39})
 \beq{45}
 \frac{\hbar}{\eta^R}=\int_0^1 K_1(\phi_x)d\phi_x=\frac{e^2}{\hbar}\int_0^1\frac{C_0^2(N_0)}{C_g^2}R_q(N_0)dN_0
 \eeq
We note that $\int_0^1\frac{C_0(N_0)}{C_g}dN_0=1$ due to the periodicity of $F(\phi_x)$.
  An extensive study \cite{guinea,golubev1,kagalovsky,bh} of $M^*(0)$  shows that it satisfies $M^*(0)>m$ and that for large $\eta$ (the bare interaction parameter) $M^*(0)/m\sim \eexp{\pi\eta}\gg 1$. Hence
  \beq{46}
  \frac{C_0}{C_g}=1-O(\eexp{-\pi\eta}) \qquad \eta\gtrsim 1
  \eeq
  and $C_0\rightarrow C_g$ for large $\eta$.

At this stage we can already propose an interesting experiment for the SEB. By analogy with $E= \hbar {\dot\phi}_x$ in the ring, we propose measuring the response to a gate voltage that is linear in time $N_0\sim t$. This leads to a DC current into the Coulomb box whose dissipation is the average in Eq. (\ref{e45}). This average is predicted to be quantized, at least for $\eta>\eta^R$, as shown below.

\section{Semiclassical RG and numerics}

\subsection{Perturbations and RG}
We study here the action (\ref{e12}) with a perturbation series for correlation functions.
Consider first the correlation $C_{t',t}=\langle \theta_{t'}\theta_t\rangle$, which to 1st order is
\beq{47}
&\ & C^{(1)}_{t,t'} =\avg{\theta_{t'}\theta_t (-S_{int})}_{S_0}
= \int_{t_1,t_2} B_{t_1,t_2} \cos v(t_1-t_2) R_{t,t_1} R_{t',t_2} \nonumber\\
\eeq
In Fourier space
\beq{48}
C^{(1)}_\omega = \abs{R_\omega}^2 B^v_\omega
\eeq
 where $ B^v_\omega =\frac12 \left( { B_{\omega+v}+B_{\omega-v} } \right)$.
 Since $C^{(1)}_{t'=t}$ is divergent it is useful to evaluate ${\tilde C}_{t,t'}=\langle[\theta_t-\theta_{t'}]^2\rangle$, which to 1st order is, with $\tau=t-t'$ ($\tau\gg 1/\omega_c$),
\beq{49}
\tilde{C}_{\tau} = \int_\omega B^v_{\omega} \abs{R_\omega}^2  (1-\cos \omega \tau) \approx \frac{2\hbar}{\pi \eta}
\left \{ \matrix{   \ln(\frac{\eta\tau}{m}) \ \ \  \tau <\frac{1}{v} \cr\cr
\half\pi v \tau \ \ \ \frac{1}{v} < \tau  } \right. .
\eeq
For $E=0$ the angular position diffuses logarithmically, while for $E\neq 0$ the long time fluctuation is linear in time.

Consider next the response function to 2nd order in $S_{int}$,
\beq{50}
&\ & R^R_{t,t'} =i\avg{\hat\theta_{t'}\theta_t}= R_{t,t'} + R^{(1)}_{t,t'} + R^{(2)}_{t,t'}
= R_{t,t'} +
i \avg{\hat\theta_{t'}\theta_t (-S_{int} +\half S_{int}^2 )}_{S_0}
\eeq
Note that the disconnected terms in the
perturbation $\avg{S_{int}^n}_{S_0} $ vanish for any order $n$, due to the normalization $Z=1$.
The first order response function is
\beq{51}
&\ & R^{(1)}_{t,t'} = -i\frac12 \int_{t_1,t_2} B_{t_1,t_2}
\avg{\hat\theta_{t_1}\hat\theta_{t_2} \cos(\theta_{t_1}-\theta_{t_2})\hat{\theta}_{t'}\theta_t}_{S_0}
\eeq
The result in frequency variable is (see Appendix B)
\beq{52}
&\ & R^{(1)}_\omega = R_\omega^2 \int_{\omega_1} R_{\omega_1} \left[B^v_{\omega_1}-B^v_{\omega-\omega_1}\right] =  R_\omega^2 \int_t R_t B_t \cos vt \ (\eexp{i\omega t}-1)
\eeq
 We note that for $v=0$ FDT is maintained, to this order,
$C^{(1)}_\omega \vert_{v=0} = \im {R_\omega} \hbar\ \sign(\omega)$.

The renormalized $\eta$  to first order is then
\beq{53}
\label{1st_order_eta}
 \frac1{\eta^R_1}=&&\lim_{\omega\rightarrow 0 } (-i\omega) R^{(1)}_\omega =
\lim_{\omega\rightarrow 0 }
\frac{-i\omega}{(-i\omega)^2 \eta^2} \int_t R_t B_t \cos v t \ (i \omega t) \nonumber\\
&&=\frac1{2\eta^2}  \ln(1+\omega_c^2/v^2)
= -\frac{\ln v/\omega_c}{\eta^2} + \order{v}
\eeq

Considering next the 2nd order in (\ref{e50}) we obtain (see Appendix B)
\beq{54}
&\ & R^{(2)}_\omega = R_\omega^2 \left( -\frac12 \int_t R_t B_t \cos v t \
(\eexp{i\omega t}-1) \ \tilde{C}_t^{(1)} +
 \int_t R^{(1)}_t B_t \cos v t \
(\eexp{i\omega t}-1) + \nonumber \right. \\
&\ & \left. R_\omega \left[\int_t R_t B_t \cos v t \
(\eexp{i\omega t1}-1) \right]^2
- \int_{t1,t2} R_{t_1} B_{t_1} B_{t_2} \sin v t_1 \sin v t_2
(1-\eexp{i \omega t_1}) t_1 \right)
\eeq

Denoting  the contribution of the last term in (\ref{e54}) as $\delta(\frac{1}{\eta_2^R})$ we obtain for the renormalized dissipation to 2nd order  (with $\ln v\rightarrow \ln v/\omega_c$ implied below)
\beq{55}
&\ & \frac1{\eta_2^R} = \frac1\eta - \frac{\ln v}{\eta^2} + \frac{\ln^2 v - \ln v}{\eta^3}  +\delta(\frac{1}{\eta_2^R})
\eeq
The contribution of the last term is peculiar and depends on the order of limits taken.
 We define a nonequilibrium limit where $\eta^R$ is evaluated for a strictly DC field, i.e. $\omega\rightarrow 0$ is taken first, and then a  logarithmically divergent $E\neq 0$ term is obtained, i.e.
\beq{56}
\label{eta_d}
&\ & \delta(\frac{1}{\eta_2^R}) = \frac1{\eta^2} \lim_{ v \rightarrow 0 }
\lim_{\omega\rightarrow 0 } \frac1{i\omega}
\int_{t1,t2} R_{t_1} B_{t_1} B_{t_2} \sin v t_1 \sin v t_2
(1-\eexp{i \omega t_1}) t_1 = \nonumber \\
&\ &  -\frac1{\eta^3} \lim_{ v \rightarrow 0 }
 \int_{t1} R_{t_1} B_{t_1} \sin v t_1 \ t_1^2
\int_{t2} R_{t_2} B_{t_2} \sin v t_2 =
\lim_{ v \rightarrow 0 } \frac1{\eta^3}
\int^{\infty} \sin(v t_1) \times \int^{\infty} \sin(v t_2)/t_2^2 = \nonumber \\
&\ & \lim_{ v \rightarrow 0 } \frac1{\eta^3} \frac1{v}
\times  v \ln v +\order{v} = \frac1{\eta^3} \ln v
\eeq
Considering next the alternative equilibrium  order of limits, i.e. first $E\rightarrow 0$, we obtain
\beq{57}
\lim_{\omega\rightarrow 0 } \lim_{ v \rightarrow 0 } \sin(v t_1) \sin(v t_2) = 0
\eeq
hence $\delta(\frac{1}{\eta_2^R})=0$.  The renormalized $\eta$ to second order is then
\beq{58}
\label{RG1}
&\ & \frac1{\eta_2^R} = \frac1\eta - \frac{\ln v}{\eta^2} + \frac{\ln^2 v + b_0 \ln v}{\eta^3}
\eeq
where $b_0$ depend on the order of limits, the nonequilibrium case has $b_0=0$, while the equilibrium one has $b_0=-1$. The latter case is in fact the known equilibrium result \cite{hofstetter}. The distinction between the two limits will become more pronounced in the full quantum treatment.

\subsection{Numerical solution of the Langevin Equation}

We solve the nonlinear Langevin equation numerically.
The time is discretized to $ t = T/N \times (1,2,...N)$, with $T$ the total time span of
system.
The noise term $\xi^i_t$ is generated numerically using a
discrete Fourier transform of $\xi^i_\omega = \sqrt{B_\omega T} \mathcal{R}^i$ where
$\mathcal{R}^i$ is a unit white Gaussian noise.
The correlation function linearity requires introducing a high frequency
cutoff $\tau_0$. We choose the cutoff to be in Lorenzian form $B_\omega = \hbar \eta {\abs{\omega}}/[{1+\omega^2\tau_0^2}]$, in the following section we explain the importance of
this choice.

We solve the equation in iterative procedure. Using the convolution form
\beq{59}
\theta_t = \int_{t'} R_{t,t'}
\left[ \xi^x_{t'} \cos \theta_{t'} - \xi^y_{t'} \sin \theta_{t'} - E  \right]
\eeq
starting with an arbitrary configuration of $\theta_t^{(0)}$
we calculate the right hand side (RHS) of (\ref{e59}) to find a new $\theta_t^{(1)}$.
We repeat the procedure $n$ times until the expression is saturated when
$\theta_t^{(n)}=\theta_t^{(n+1)}$.
This procedure is improved if instead of taking the convolution result
as the next order $\theta_t$ we use some mixing of that result
and of the previous $\theta_t$ configuration in the form
$\theta_t^{(m)}= (1 - \beta) \theta_t^{(m-1)} + \beta \times \mbox{RHS}$
where $\beta$ is mixing parameter.
Typically $n$ would be in order of $10^5$ and $\beta=0.1$.

  With this choice the Langevin equation takes the following form
\beq{60}
&& m \ddot{\theta}_t = \xi^x_t \cos \theta_t + \xi^y_t \sin \theta_t + E + \Delta_t \\
&& \Delta_t = \frac{\eta}{\tau_0^2} \int_{-\infty}^t
\sin[\theta_t - \theta_{t'}] \eexp{-(t-t')/\tau_0} dt' \nonumber,
\eeq
where $\Delta_t$ is a correction term defined
by  $\delta R^{-1}_\omega$ in the response function  Eq. (\ref{e19})
as $\int_{t'}\delta R^{-1}_{t,t'} [\xi^x_{t'}
\cos \theta_{t'} + \xi^y_{t'} \sin \theta_{t'} + E]  =
- \int_\omega m\omega^2 \Delta_\omega$

In the numerical system we have now four time scales, two numerical time scales, i.e.
 the time segment $\Delta \tau = {\bar T}/N$ and the time span  ${\bar T}$ , as well as the two physical
high frequency cutoffs, $1/\tau_0$  for the noise and $\omega_c$ the mass cutoff.
The region of interest corresponds to velocity
$v^R=\avg{\dot{\theta}_t}$ between the time scales
$ \Delta \tau  \ll \tau_0 < 1/\omega_c \ll 1/v^R \sim 1/v < {\bar T}$.
The inequality $\tau_0 < 1/\omega_c$ is useful since
we compare the numerical result to an asymptotic result in which
$\omega_c$ rather than $1/\tau_0$ is the high frequency cutoff.

With the result for $\theta_t$
we can find the renormalized $1/{\eta^R} = dv^R/dE$ with
$v^R=\avg{\dot{\theta}_t}$ where the average $\avg{...}$ reflects an
average on both the time domain $t>1/\omega_c$
and on numerous realizations of the noise.

In the left panel of Fig.\ref{eta_R_fig} our numerical solution for the Langevin equation is shown, including a fit to the second order with $b_0=0$. On the right panel the 1st order is subtracted with either the nonequilibrium  $b_0=0$ or the equilibrium $b_0=-1$. The first is in fact a better fit for the numerical data.
When $1/v$ approaches the simulation time span ${\bar T}$
the numerics become unreliable, as the particle cannot complete even one revolution in time ${\bar T}$;
a plateau is then observed at low $E$.

\begin{figure}[h]
\begin{center}
\includegraphics[scale=0.4]{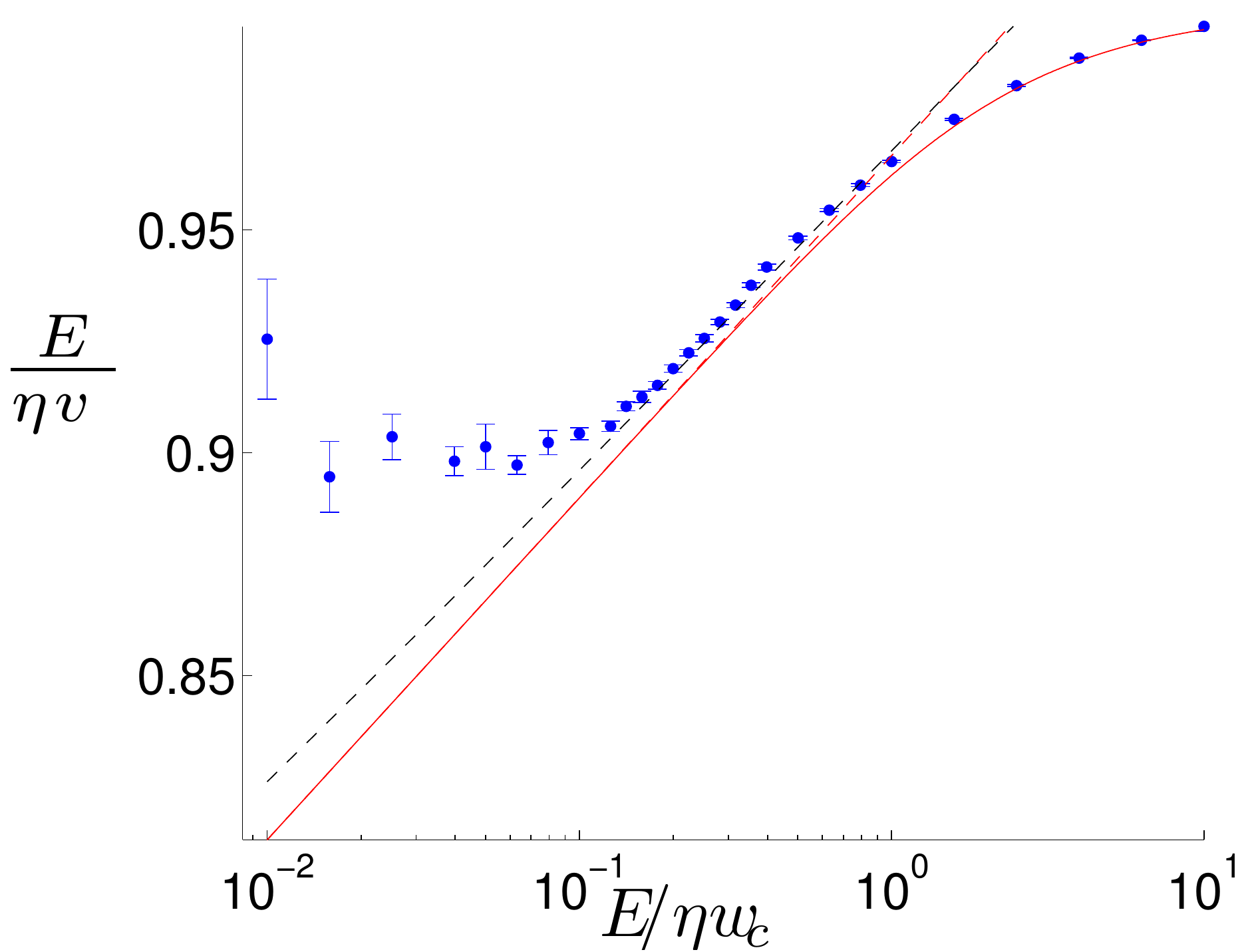}
\includegraphics[scale=0.4]{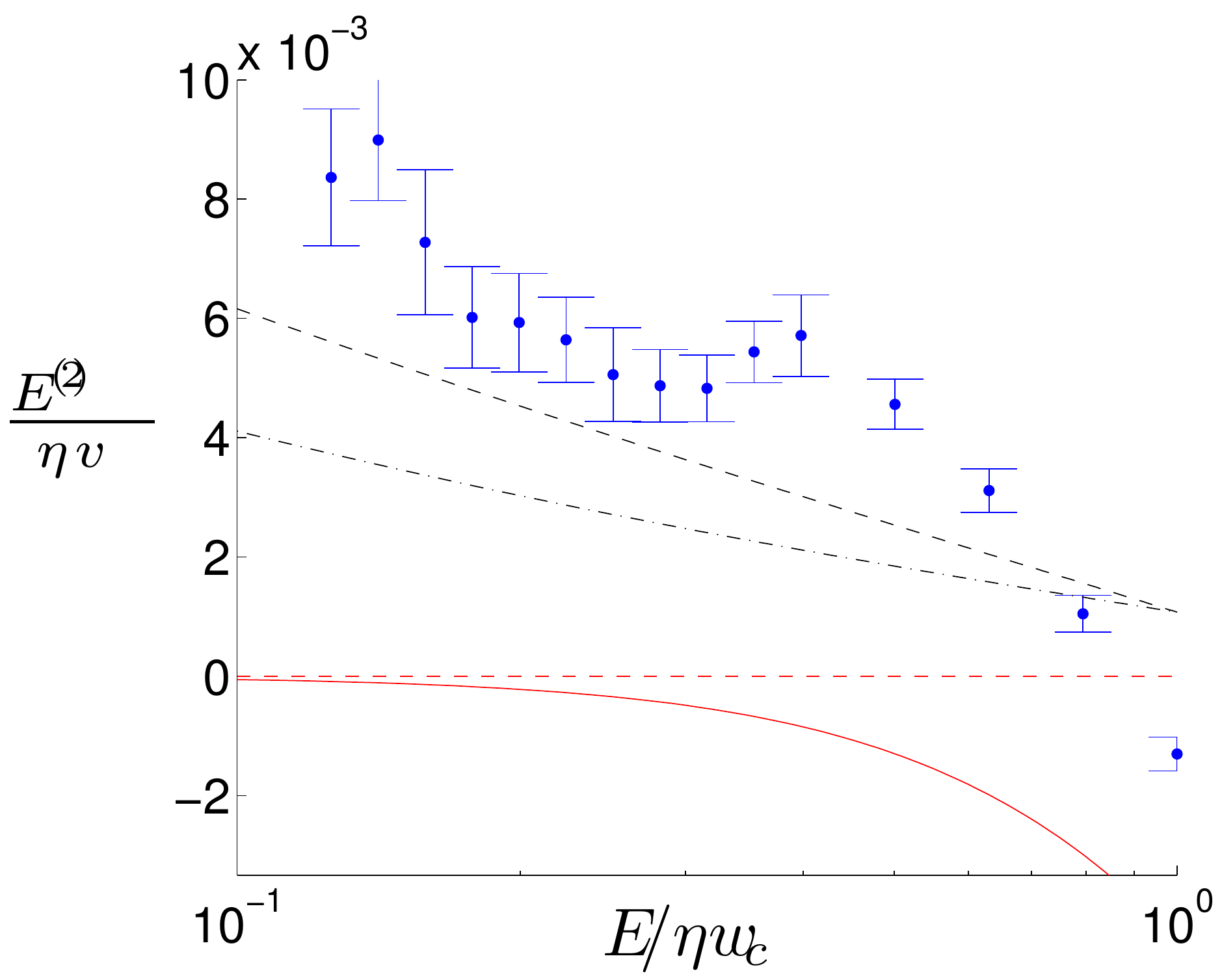}
\end{center}
\caption{
{\bf Left panel}: Velocity-field relation for Eq. (\ref{e60})
with $\eta=30\hbar/\pi$,
$\omega_c = 100/\tau_0$ and $\tau_0=20 \Delta\tau$. Here $N=2^{15}$, $\Delta\tau = 1/20$.
The circles are numerical data, the full red line is a 1st order
perturbation in $1/\eta$, the dashed lower red line is its logarithmic expansion for large
$\ln v/\omega_c$ and the dashed upper (black) line includes the 2nd order logarithmic term, corresponding to Eq. (\ref{RG1}) for $b_0=0$. Note that the data is not reliable for $E/\eta\omega_c \lesssim 1/(\Delta\tau N \omega_c) \sim 0.06$.
\newline {\bf Right panel}: The same data and line types after subtracting the 1st order terms, i.e. $\frac{E^{(2)}}{\eta v}= \frac{E}{\eta v}-1-\frac{\hbar}{\pi\eta}(\ln \frac{v}{\omega_c}-1)$. An additional dash-dotted line corresponds to $b_0=-1$, which is a worse fit to the data than $b_0=0$   (dashed upper line). Note that the numerical data displays $E/v$ rather than $dE/dv$, hence Eq. (\ref{e53}) acquires a $-1$ term.}
\label{eta_R_fig}
\end{figure}


With the numerical results for $\theta_\tau$ we can also generate the correlation function
$\tilde{C}_\tau = \avg{[\theta_\tau -\theta_0]^2 }$, the first order
perturbation for this correlation function is given in Eq. (\ref{e49}).
In Fig. \ref{Fig:tildeC} we plot this correlation function as a function of the
time separation $\tau$ for the same parameters as in Fig.\ref{eta_R_fig},
with and without a finite field.
The data is fairly close to the 1st order result (\ref{e49}) for not too long times, i.e. for zero field the correlation has a subdiffusion
logarithmic behavior while for finite force the correlation has a diffusion ($\sim\tau$) behavior.

\begin{figure}[h]
\begin{center}
\includegraphics[scale=0.4]{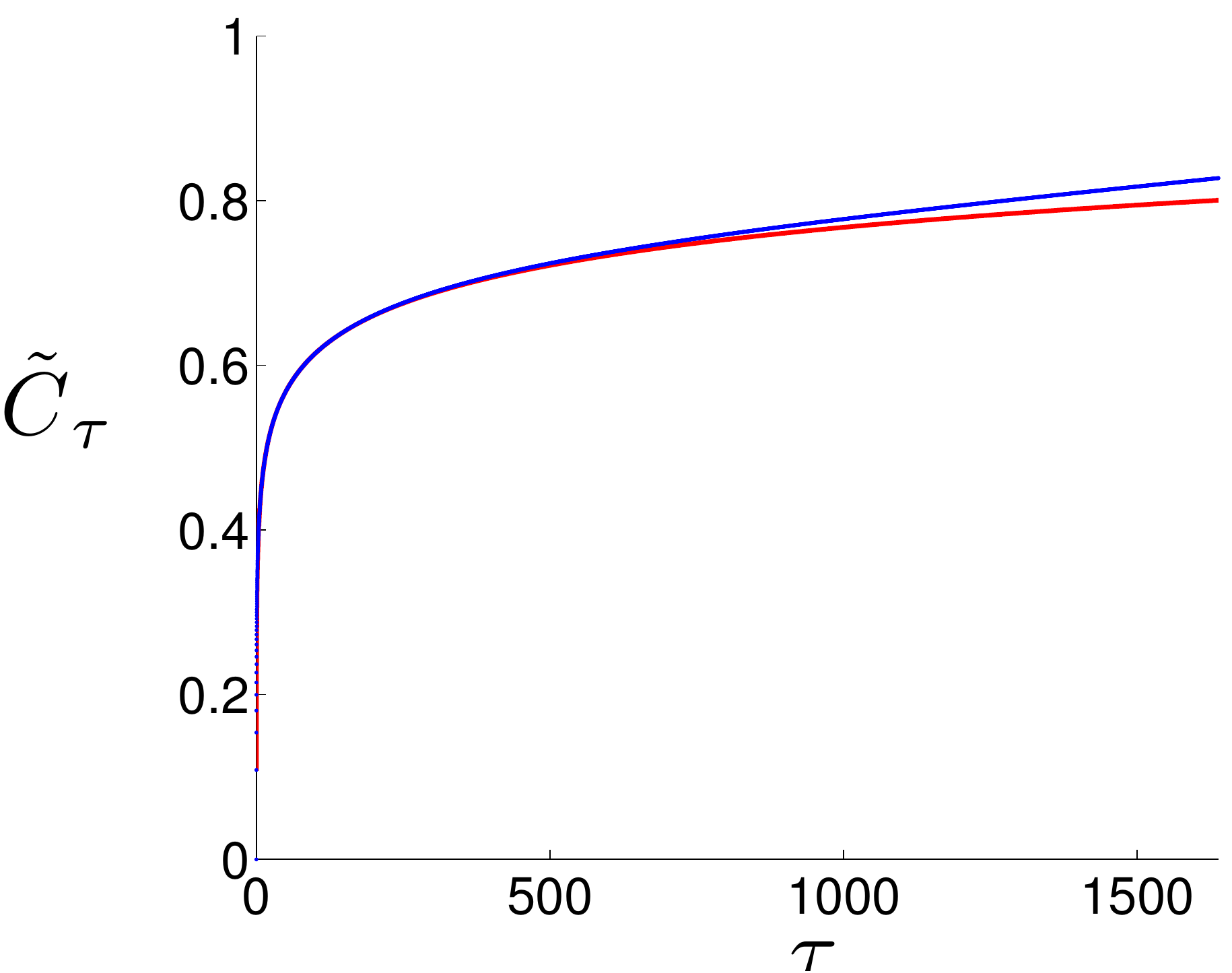}
\includegraphics[scale=0.4]{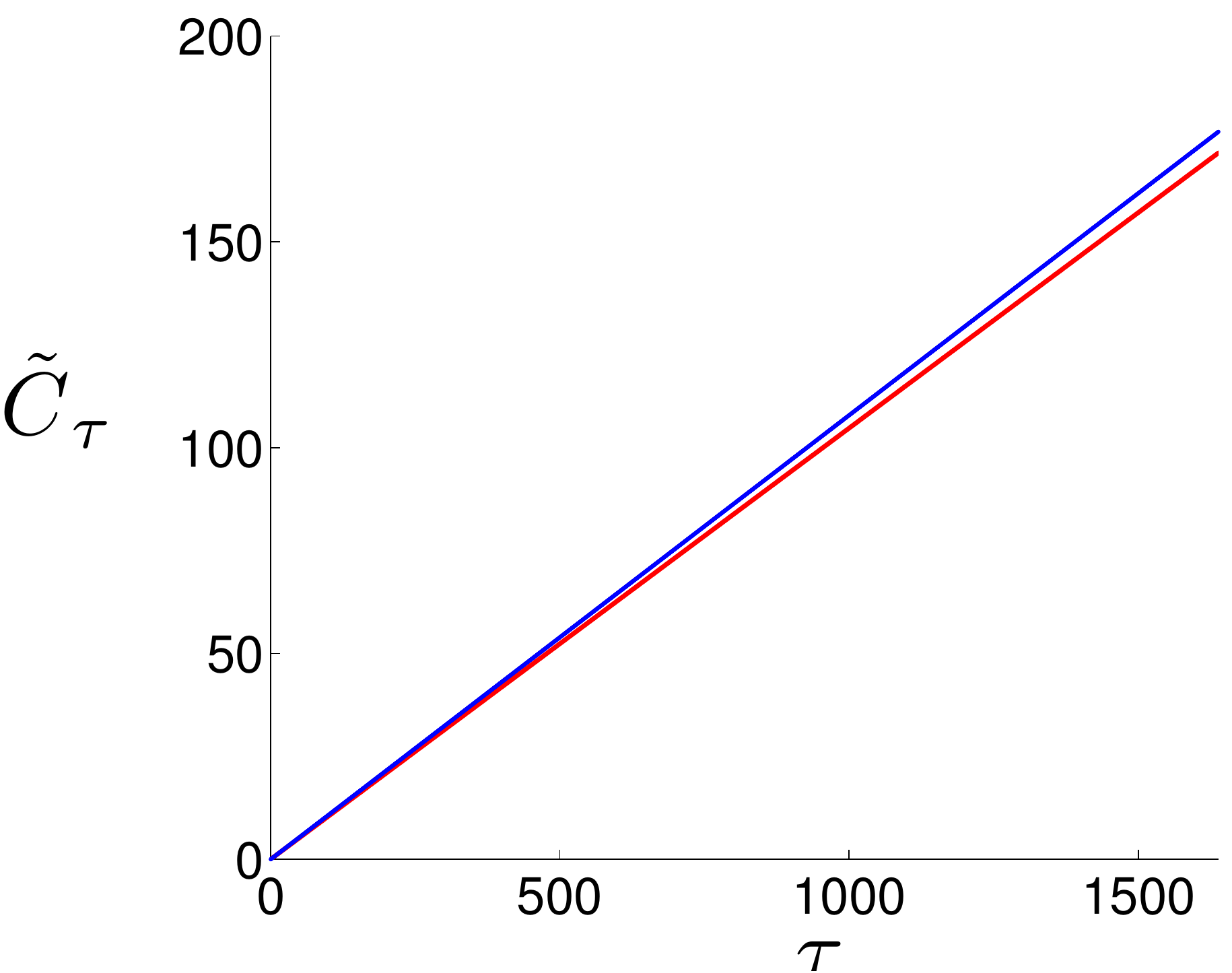}
\end{center}
\caption{
{\bf Left panel}: The correlation function  $\tilde C_{\tau}$ as a function of time (Blue)
and the asymptotic results of Eq. (\ref{e49}) (red) for $E=0$.
{\bf Right panel}: The correlation function as a function of time (Blue) and the asymptotic
results of Eq. (\ref{e49}) for $E/\eta = 1$ and $\tau_0=1$.
}\label{Fig:tildeC}
\end{figure}

\section{Quantum RG}

\subsection{Perturbations from $S_{int}$}

Consider now the definition $\eta^R$ in Eqs. (\ref{e28},\ref{e29})
\beq{61}
 -i\delta E^{(1)}&=&\bra \frac{\delta
S_{int}}{\delta {\hat \theta}_t}\ket_0=
\frac{2}{\hbar}\int_{t'}B_{t,t'}\bra
\cos(\frac{\hbar}{2}{\hat \theta_t})\sin(\frac{\hbar}{2}{\hat
\theta}_{t'})\cos(vt-vt'+\delta\theta_{t}-\delta\theta_{t'})\ket_0=\nonumber\\
&&\frac{2}{\hbar}\int_{t'}B_{t,t'}\sum_{\sigma,\sigma',\mu=\pm}\frac{\sigma'}{8i}\bra\eexp{\half
i\hbar\sigma{\hat \theta}_t+\half i\hbar\sigma'{\hat
\theta}_{t'}+i\mu(vt-vt'+\delta\theta_{t}-\delta\theta_{t'})}\ket_0=\nonumber\\
&&\frac{2}{\hbar}\int_{t'}B_{t,t'}\sum_{\sigma,\sigma',\mu=\pm}\frac{\sigma'}{8i}\eexp{-\half\mu\hbar
(\sigma iR_{t't}-\sigma' iR_{tt'})+i\mu(vt-vt')} \eeq
For $t<t'$ the term $\sigma' R_{tt'}=0$ and then $\sum \sigma'=0$.
The result is then finite only for $t>t'$; defining
$\mu'=\mu\sigma'$,
\beq{62}&&=\frac{2}{\hbar}\int_{t'}B_{t,t'}\sum_{\sigma',\mu'=\pm}\frac{\sigma'}{4i}
\eexp{i\sigma'\mu'(vt-vt')+\half i\hbar \mu' R_{tt'}}=
i\frac{2}{\hbar}\int_{t'}B_{t,t'}\sin
v(t-t')\sin(\half\hbar R_{tt'}) \eeq
Hence the force correction is
\beq{63} \label{eq63}
\delta E^{(1)}=-\frac{2}{\hbar}\int_{\tau}B_\tau\sin(\half\hbar R_{\tau})\sin
(v\tau) \eeq
so that using Eq. (\ref{e29}) and performing the calculation of the integrals
with arbitrary cutoffs $\tau_0$ and $\omega_c^{-1}=m/\eta$ one obtains:
\beq{64} \label{eq64}
\frac{1}{\eta^R}=\frac{1}{\eta}-\frac{2}{\pi\eta}[\sin (\frac{\hbar}{2\eta}) \ln (v/\omega_c) + C + O(1/v)]
\eeq
where the constant $C$ depends on $\tau_0$ and $\omega_c$. Although we will not
need it below, its detailed form is given in the Appendix \ref{app:integral} in the limit
$\tau_0=0$.

Consider next 2nd order in $S_{int}$,

\beq{65}
&&i\delta E^{(2)}=\half\bra \frac{\delta}{\delta {\hat\theta}_{t_1}}S_{int}^2\ket=
\half 4 (\frac{2}{\hbar^2})^2\frac{\hbar i}{2\cdot 2^6}\sum_{\epsilon_i\sigma,\sigma'=\pm}
\epsilon_2\epsilon_3\epsilon_4\int_{t_2,t_3,t_4}B_{t_1,t_2}B_{t_3,t_4}
\eexp{i\sigma v(t_1-t_2)+i\sigma'v(t_3-t_4)}\nonumber\\
&&\times \bra \eexp{\half i(\epsilon_1{\hat\theta}_{t_1}
+\epsilon_2{\hat\theta}_{t_2}+\epsilon_3{\hat\theta}_{t_3}+\epsilon_4{\hat\theta}_{t_4})
+i\sigma(\theta_{t_1}-\theta_{t_2})+i\sigma'(\theta_{t_3}-\theta_{t_4})}\ket_0
\eeq
Note that $\delta/\delta{\hat\theta}_{t_1}$ can be applied also at either $t_2,t_3,t_4$ and all these terms are identical since $\sin (\half\hbar{\hat\theta}_{t_i})$ appears in the same form for all $t_i$, hence a factor 4. Now change all $\epsilon_i,\sigma,\sigma'\rightarrow -(\epsilon_i,\sigma,\sigma')$ and define $\sigma'=\sigma\mu$ to obtain
\beq{66}
&&i\delta E^{(2)}=\frac{i^2}{16\hbar^3}\sum_{\epsilon_i\sigma,\mu=\pm}
\epsilon_2\epsilon_3\epsilon_4\sigma\int_{t_2,t_3,t_4}B_{t_1,t_2}B_{t_3,t_4}
\sin [v(t_1-t_2)+\mu v(t_3-t_4)]\nonumber\\
&&\times \exp \{-\half\hbar\bra \sigma (\epsilon_1{\hat\theta}_{t_1}
+\epsilon_2{\hat\theta}_{t_2}+\epsilon_3{\hat\theta}_{t_3}+\epsilon_4{\hat\theta}_{t_4})
[\theta_{t_1}-\theta_{t_2}+i\mu(\theta_{t_3}-\theta_{t_4})]\ket_0\}\nonumber\\
&&=\frac{-1}{8\hbar^3}\sum_{\epsilon_i,\mu=\pm}\epsilon_2\epsilon_3\epsilon_4
\int_{t_2,t_3,t_4}B_{t_1,t_2}B_{t_3,t_4}A_2\sin[v(t_1-t_2+\mu v(t_3-t_4)]
\eeq
where
\beq{67}
A_2=&&\exp\{\half i\hbar\epsilon_1(-R_{t_2,t_1}+\mu R_{t_3,t_1}-\mu R_{t_4,t_1})\}\times\nonumber\\
&&\exp\{\half i\hbar\epsilon_2(R_{t_1,t_2}+\mu R_{t_3,t_2}-\mu R_{t_4,t_2})\}\times\nonumber\\
&&\exp\{\half i\hbar\epsilon_3(R_{t_1,t_3}-R_{t_2,t_3}-\mu R_{t_4,t_3})\}\times\nonumber\\
&&\exp\{\half i\hbar\epsilon_4(R_{t_1,t_4}-R_{t_2,t_4}+\mu R_{t_3,t_4})\}\,.
\eeq

 Note that in $A_2$ if $t_2$ is the maximal time then its second factor =1 and $\sum_{\epsilon_2}\epsilon_2=0$. similarly, if $t_3$ (or $t_4$) is the maximal time, the the 3rd (or 4th) factor =1 and $\sum_{\epsilon_3}\epsilon_2=0$ (or $\sum_{\epsilon_4}\epsilon_2=0$). Therefore $t_1$ must be the maximal time and the 1st factor =1. The result is symmetric in $t_3\leftrightarrow t_4$, so choose $t_3>t_4$, with factor 2. Hence 3 time orderings, denoted by A,B,C,
$\delta E^{(2)}=\delta E_A+\delta E_B+\delta E_C$,
\beq{68}
\delta E_A=&&\frac{4}{\hbar^3}\sum_{\mu}\int_{t_1>t_2>t_3>t_4}\sin (\half\hbar R_{t_1,t_2})\sin [\half\hbar(R_{t_1,t_3}-R_{t_2,t_3})]\sin[\half\hbar(R_{t_1,t_4}-R_{t_2,t_4}+\mu R_{t_3,t_4})]\nonumber\\
&&\times B_{t_1,t_2}B_{t_3,t_4}\sin[v(t_1-t_2)+\mu v(t_3-t_4)]\nonumber\\
\delta E_B=&&\frac{4}{\hbar^3}\sum_{\mu}\int_{t_1>t_3>t_2>t_4}\sin [\half\hbar(R_{t_1,t_2}+\mu R_{t_3,t_2})]\sin (\half\hbar R_{t_1,t_3})\sin[\half\hbar(R_{t_1,t_4}-R_{t_2,t_4}+\mu R_{t_3,t_4})]\nonumber\\
&&\times B_{t_1,t_2}B_{t_3,t_4}\sin[v(t_1-t_2)+\mu v(t_3-t_4)]\nonumber\\
\delta E_C=&&\frac{4}{\hbar^3}\sum_{\mu}\int_{t_1>t_3>t_4>t_2}\sin[\half\hbar(R_{t_1,t_2}+\mu R_{t_3,t_2}-\mu R_{t_4,t_2})]\sin (\half\hbar R_{t_1,t_3})\sin [\half\hbar(R_{t_1,t_4}+\mu R_{t_3,t_4})]\nonumber\\
&&\times B_{t_1,t_2}B_{t_3,t_4}\sin[v(t_1-t_2)+\mu v(t_3-t_4)]
\eeq
B and C terms can be time ordered as A by $t_2\leftrightarrow t_3$ in B and $t_2\rightarrow t_4,
t_4\rightarrow t_3,t_3\leftrightarrow t_2$ in C. In terms of the $\mu=\pm$ components,
\beq{69}
\delta E_A^++\delta E_C^-&=&\frac{4}{\hbar^3}\int_A \sin (\half\hbar R_{t_1,t_2})\sin [\half\hbar(R_{t_1,t_3}-R_{t_2,t_3})]\sin[\half\hbar(R_{t_1,t_4}-R_{t_2,t_4}+ R_{t_3,t_4})]\nonumber\\
&&\times [B_{t_1,t_2}B_{t_3,t_4}+B_{t_1,t_4}B_{t_2,t_3}]\sin[v(t_1-t_2+t_3-t_4)]\nonumber\\
\delta E_A^-+\delta E_B^-&=&\frac{4}{\hbar^3}\int_A \sin (\half\hbar R_{t_1,t_2})\sin [\half\hbar(R_{t_1,t_3}-R_{t_2,t_3})]\sin[\half\hbar(R_{t_1,t_4}-R_{t_2,t_4}- R_{t_3,t_4})]\nonumber\\
&&\times [B_{t_1,t_2}B_{t_3,t_4}+B_{t_1,t_3}B_{t_2,t_4}]\sin[v(t_1-t_2+t_4-t_3)]\nonumber\\
\delta E_B^++\delta E_C^+&=&\frac{4}{\hbar^3}\int_A \sin (\half\hbar R_{t_1,t_2})\sin [\half\hbar(R_{t_1,t_3}+R_{t_2,t_3})]\sin[\half\hbar(R_{t_1,t_4}-R_{t_3,t_4}+ R_{t_2,t_4})]\nonumber\\
&&\times [B_{t_1,t_3}B_{t_2,t_4}+B_{t_1,t_4}B_{t_2,t_3}]\sin[v(t_1-t_3+t_2-t_4)]
\eeq

In appendix E we derive the $\ln^2 v$ coefficient directly for the single cutoff case where $\tau_0=0$. Here we proceed with a shorter indirect method. In general we have two cutoffs $m/\eta,\tau_0$ in Eq. (\ref{e07}) and we define $\tau_1(m/\eta,\tau_0)$ as the cutoff time for the response $R_t$, Eq. (\ref{e07}).
For the purpose of identifying the leading $\ln^2 v$ term we take a formal limit such that this cutoff time is $\tau_1\rightarrow 0$. We will eventually restore physical cutoffs corresponding to $m/\eta, \tau_0$ in $R_{t}$.
The only cutoff for now is $\tau_0$ in $B(\omega)$, Eq. (\ref{e06}). In this limit $R_{t}\rightarrow \frac{1}{\eta}\Theta(t)\eexp{-\delta t}$ where $\delta\rightarrow +0$ to ensure the retarded nature (poles of $1/(\omega+i\delta)$). The significant virtue of this limit is that the 1st two equations of (\ref{e69}) vanish since $R_{t_1,t_3}-R_{t_2,t_3}\rightarrow 0$, leaving just the last form. The evaluation of $\delta E^{(2)}$ in this limit is straightforward (Appendix D), leading to
\beq{70}
\delta E^{(2)}=\frac{4\eta^2}{\pi^2\hbar}\sin^2(\frac{\hbar}{2\eta})\sin(\frac{\hbar}{\eta})\cdot
v\ln(v\tau_0)[\ln(v\tau_0)+1]
\eeq
Hence from (\ref{e29})
\beq{71}
\frac{1}{\eta^{R(2)}}=\frac{4}{\pi^2\hbar}\sin^2(\frac{\hbar}{2\eta})\sin(\frac{\hbar}{\eta})\cdot
[\ln^2(v\tau_0)+3\ln(v\tau_0)+1]
\eeq

So far $\delta E^{(2)}$ is calculated in a formal limit $\tau_1\rightarrow 0$. We proceed by asserting that for any $\tau_0,\tau_1$ the leading singularity as $v\rightarrow 0$  is a $\ln^2v$ term, as expected for a 2-loop calculation. This term must involve an $\eta$ dependent function $f_{\eta}(\tau_0,\tau_1)$ that has dimensions of time. Fixing the coefficient of $\ln^2 [vf_{\eta}(\tau_0,\tau_1)]$ as in Eq. (\ref{e71}), we have $f_{\eta}(\tau_0,0)=\tau_0$ while for $\tau_0\rightarrow 0$, when $\tau_1\rightarrow m/\eta=1/\omega_c$ we must have the form $f_{\eta}(0,\tau_1)=b(\eta)\tau_1=b(\eta)/\omega_c$. The 2-loop correction Eq. (\ref{e71}) becomes at $\tau_0=0$
\beq{72}
\frac{1}{\eta^{R(2)}}=\frac{4}{\pi^2\hbar}\sin^2(\frac{\hbar}{2\eta})\sin(\frac{\hbar}{\eta})\cdot
\ln^2[\frac{v}{\omega_c}b(\eta)] +O(\ln v)
\eeq
The renormalized friction has therefore the form
\beq{73}
\frac{1}{\eta^R}&=&\frac{1}{\eta}-\frac{2}{\pi\eta}\sin(\frac{\hbar}{2\eta})\ln[\frac{v}{\omega_c}]+
\frac{4}{\pi^2\hbar}\sin^2(\frac{\hbar}{2\eta})\sin(\frac{\hbar}{\eta})
\{\ln^2[\frac{v}{\omega_c}]+b_0(\eta)\ln[\frac{v}{\omega_c}]\}
\eeq
We have thus identified the coefficient of the $\ln^2$ term; this coefficient is also identified by the more lengthy calculation of the $\tau_0=0$ case in appendix E. In appendix E we further show that the coefficient of the $\ln v$ term, i.e. $\sin^2(\frac{\hbar}{2\eta})\sin(\frac{\hbar}{\eta})b_0(\eta)$, has at least one factor of
$\sin(\frac{\hbar}{2\eta})$. Hence the perturbation expansion as well as the following RG analysis are justified near the zeroes of $\sin(\frac{\hbar}{2\eta})$.

We note that in the semiclassical limit the perturbation expansion is in $R^{2n-1}B^n/\eta^2\sim 1/\eta^{n+1}$ for large $\eta$; in the quantum case the $R^{2n-1}$ factors become periodic functions. The main conclusion is that there is a new small parameter in the perturbation series, $\sin(\frac{\hbar}{2\eta})$.

\subsection{Perturbations from $S_c$}
Here we consider the $S_c$ interaction in Eq. (\ref{e21}). The $S_c$ terms are
\beq{74}
\langle {\hat\theta}_{t'}\theta_t S_c\rangle=\langle {\hat\theta}_{t'}\theta_t S_c^2\rangle=0
\eeq
However, the mixed term and the corresponding correction to $1/\eta$ are
\beq{75}
\delta R^m_{t,t'}&=&i\langle {\hat\theta}_{t'}\theta_t S_c S_{int}\rangle  \nonumber\\
\Rightarrow \qquad \frac{1}{\eta^m}&=&\frac{2}{\pi\hbar}[\sin\frac{\hbar}{2\eta}(\sin\frac{\hbar}{\eta}-\frac{\hbar}{\eta})
+\frac{\hbar}{2\eta}\cos\frac{\hbar}{2\eta}(\sin \frac{\hbar}{\eta}-\frac{\hbar}{\eta})]\ln(v\tau_1)
\eeq
 which does not vanish at $\sin\frac{\hbar}{2\eta}=0$. Note, however that this term is $\sim \hbar^3$, i.e. a 3 -loop term. Furthermore, other response functions do show such zeroes. E.g. for the ${\bar R}_{t,t'}$ correlation (Eq. (\ref{e77}) below) we have $\langle \theta_t\sin\frac{\hbar}{2}{\hat\theta}_{t'} S_c\rangle=0$ to 1st order, while in 2nd order
\beq{76}
\delta {\bar R}^m_{t,t'}&=&\frac{2i}{\hbar}\langle \theta_t\sin\frac{\hbar}{2}{\hat\theta}_{t'}S_cS_{int}\rangle\nonumber\\
\Rightarrow \frac{1}{{\bar\eta}^m}&=&\frac{2}{\pi\hbar}\sin\frac{\hbar}{\eta}(\sin \frac{\hbar}{\eta}-\frac{\hbar}{\eta})\ln(v\tau_1)
\eeq
 We note that there are many other operators that have vanishing perturbations at $\sin\frac{\hbar}{2\eta}=0$ to 2nd order in $S_{int},S_c$, e.g. the dissipation term in Eq. (\ref{e09}) $\bra \theta_t\sin(\hbar{\hat\theta}_{t'})\ket$, or the response to an AC field with frequency $v$ $\langle \theta_t\cos\delta\theta_{t'}\sin\frac{\hbar}{2}{\hat\theta}_{t'}\rangle$.

\subsection{RG analysis}

We note that in (\ref{e73}) $g=\frac{2}{\pi}\sin(\frac{\hbar}{2\eta})$ acts as an unexpected small parameter for the expansion, since all
divergences vanish when $g=0$. It raises the interesting possibility that $g=0$ be viewed as a RG fixed point.
For that we need to find a renormalized coupling which obeys multiplicative RG, the simplest choice being
$g_R=\frac{2}{\pi}\sin(\frac{\hbar}{2\eta^R(E)})$. The question is then whether the $\beta$-function $\beta=- E \partial_E g_R$ can be written only in terms of $g_R$. Although the non-periodic $1/\eta$ factor in (\ref{e73}) appears at first problematic, we propose that resummation from higher loops, which allows for higher order terms $O(\frac{1}{\eta^4})$ changes the 1-loop term in (\ref{e73}) by $\frac{\hbar}{2\eta}\rightarrow\sin(\frac{\hbar}{2\eta})$.

 To further motivate this proposal we consider the response
 \beq{77}
{\bar R}_{t,t'}=i\frac{2}{\hbar}\bra \theta_t\sin(\frac{\hbar}{2}{\hat\theta}_{t'})\ket.
 \eeq
 Physically, $\eexp{\pm i \frac{\hbar}{2}{\hat\theta}_{t'}}$  corresponds to an electric field pulse
$\delta E(t)=\pm\frac{\hbar}{2}\delta (t-t')$ or equivalently a rapid change of flux by $\pm\half$, therefore
${\bar R}_{t,t'}$ corresponds to the difference in response to these two flux pulses. Defining the dissipation parameter $\bar\eta^R$ for ${\bar R}_{t,t'}$ as in Eq. (\ref{e25}) we obtain that the 1-loop term is fully periodic with
\beq{78}
\frac{\hbar}{2\bar\eta^R}&=&\frac{\hbar}{2\eta}-\frac{2}{\pi}\sin^2(\frac{\hbar}{2\eta})\ln[\tau_1v]
\eeq
hence $\frac{\hbar}{2\eta}\rightarrow\sin(\frac{\hbar}{2\eta})$ in Eq. (\ref{e73}).

We propose then that an RG consistent theory corresponds to
\beq{79}
\frac{\hbar}{2\eta^R}&=&\frac{\hbar}{2\eta}-\frac{2}{\pi}\sin^2(\frac{\hbar}{2\eta})\ln[\tau_1v]+
\frac{4}{\pi^2}\sin^3(\frac{\hbar}{2\eta})\cos(\frac{\hbar}{2\eta})
\{\ln^2[\tau_1v]+b_0(\eta)\ln[\tau_1v]\}\nonumber\\
\eeq
Taking a sine of both sides it yields to order $g^3$, with $b_0=b_0(g=0)$,
\beq{80}
g_R=g\mp g^2\ln(v/\omega_c)+g^3[\ln^2(v/\omega_c)+b_0\ln(v/\omega_c)]
\eeq
where $\pm$ refers to $g=0$ with $\cos(\frac{\hbar}{2\eta})=\pm 1$, leading to
\beq{81}
\beta(g_R)=\frac{dg^R}{-d\ln v}=\pm g_R^2 - b_0 g_R^3+ O(g_R^4).
\eeq
This RG equation is satisfied for both $\pm$ fixed points as seen by substituting (\ref{e80}). We propose then that $g^R=0$ are exact zeroes of the perturbation expansion and the additional requirement of an RG structure leads to the result (\ref{e80}).

 Eq. (\ref{e80}) yields fixed points at $\frac{\hbar}{2\eta_n}=n\pi$ with $n=1,2,3,...$ that are attractive at $\eta>\eta_n$ and repulsive at $\eta<\eta_n$, i.e. the flow of $\eta\neq\eta_n$ is always to smaller $\eta$. At these fixed points a Gaussian evaluation yields the correlation $\langle\cos\theta_t\cos\theta_0\rangle\sim t^{-2n}$. We recall now a theorem for the lattice model \cite{spohn} where the equilibrium action with mass related cutoff is replaced by an action on a lattice resulting in an XY model with long range interactions. The theorem states \cite{spohn} that $\langle\cos\theta_t\cos\theta_0\rangle\sim 1/t^2$; this result was also derived \cite{bh} in first order in $\eta$. The range $\eta>\eta_1$ has an RG flow to $\eta_1$ and is therefore consistent with the theorem. The hypothesis of Gaussian fixed points corresponding to $n \geq 2$ is inconsistent with the theorem, i.e. $\langle\cos\theta_t\cos\theta_0\rangle$ becomes a relevant operator at the $n\leq 2$ points rendering them unstable.
Note that in the SEB problem $\cos\theta_t$ corresponds to a lead-dot voltage and its correlations determine the SET conductance \cite{AES,schon,burmistrov1}, while in the ring problem it corresponds to fluctuations in the circular asymmetry.

For $\eta < \eta_1$ the system could have
non-gaussian fixed points or a line of fixed points as hinted by the small $\eta$ perturbation \cite{bh}. The equilibrium $K_1(\phi_x)$ was evaluated for small $\eta$ and for $T\rightarrow 0$ has the form $K_1(\phi_x)\sim \delta(\phi_x-\half)/T$, i.e. the dissipation is concentrated at the single point $\phi_x=\half$. This implies from Eq. (\ref{e39}) that $\eta^R\sim T$ and thefore vanishes at temperature $T=0$. It is not clear, however, that $\eta=0$ is a fixed point in the RG sense and if so what is its range of attraction. An $\eta=0$ fixed point would imply the implausible result that the ring conductance diverges for small but finite $\eta$. We therefore expect that $\eta_1\equiv \eta^R$ is the single fixed point in this system, as illustrated in Fig. 4.

\begin{figure}[t]
\begin{center}
\includegraphics[scale=0.6]{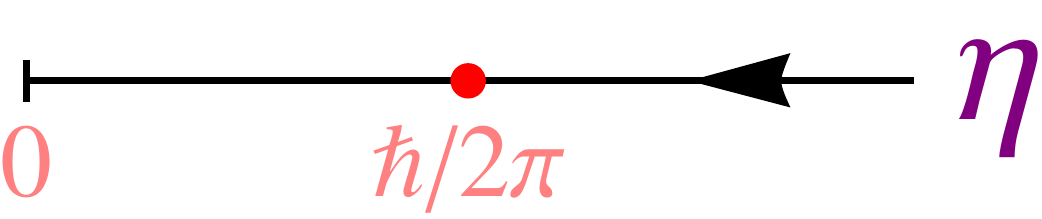}
\end{center}
\caption{RG flow of $\eta$.}
\end{figure}

\section{Discussion}

The special value $\eta^R=\hbar/(2 \pi)$ has a topological interpretation as a Thouless charge pump \cite{thouless}, as shown in the introduction. Hence a slow change in $\phi_x$ by one unit results in transporting a unit charge once around the ring if $\eta^R=\hbar/(2 \pi)$. Such quantization has been shown for cases where the spectrum has a gap \cite{thouless}, though quantized charge transport was shown also in cases without a gap \cite{aleiner,sharma}; in our case the gap vanishes \cite{falci} at flux $\phi_x=\half$. Vanishing of this gap is essential in solving for the dissipation problem in the ring via Landau-Zener transitions, as studied in related models \cite{shimshoni}.
 We note that the quantized $\eta^R$ also results from arguing that there should be a unique frequency $\omega_E =\frac{2\pi}{\hbar}E= v$ as $E \to 0$ (see discussion below Eq. \ref{e38}), as suggested by linear response.

We conclude from (\ref{e45}) that for $\eta>\eta_1\equiv\eta^R$ the SEB satisfies the quantization (see definitions in section IIE)
\beq{82}
\int_0^1\frac{C_0^2(N_0)}{C_g^2}R_q(N_0)dN_0=\frac{h}{e^2}\,.
\eeq
In particular, when $\eta/\hbar \gtrsim 1$ we have \cite{guinea,golubev1,kagalovsky,bh}  from the known $M^*/m\sim\eexp{\pi\eta/\hbar}$ and from Eq. (\ref{e06}) $C_0/C_g=1+O(\eexp{-\pi\eta/\hbar})$. We expect $R_q$ to be independent of $N_0$ at large $\eta$, hence
\beq{83}
R_q=\frac{h}{e^2}[1+O(\eexp{-\pi\eta/\hbar})]
\eeq
similar to the $N_c=1$ case \cite{mora}.

The conductance of the ring can be defined by the voltage around the ring  $2\pi E/e$ and the current  $e\langle\dot\theta\rangle/2\pi$, hence we predict that the conductance for $\eta>\eta^R$ is
 \beq{84}
 G_{ring}=\frac{e^2}{4\pi^2\eta^R} =\frac{e^2}{h}\,.
 \eeq
While this well known quantum conductance seems natural, we emphasize that it is due to the inherent nonequilibrium nature of the driving force and the specific limiting procedure of taking a DC limit before the linear response limit, Eq. (\ref{e39}).

 Finally, we consider the conditions for our proposed box experiment. The Coulomb box, i.e. a metallic quantum dot, should be connected to the electrode with $N_c\gg 1$ degenerate channels; in fact $N_c$ can be fairly small and yet reproduce the $N_c\rightarrow\infty$ case, except at exponentially small temperatures \cite{zarand}.
 By analogy with $E= \hbar {\dot\phi}_x$ in the ring, we propose measuring the response to a gate voltage that is linear in time $N_0=E t$. This leads to a DC current into the Coulomb box whose dissipation is the average in Eq. (\ref{e45}).
 The field $E$ should be sufficiently small so that $g_R$ is sufficiently near the fixed point. For an initial $g\approx 1$ integration of
 $\partial g_R/\partial \ln E=g_R^2$ yields $g_R=1/\ln (\hbar \omega_c/E) \ll g$. E.g. for $g_R\lesssim 0.1$ and a typical $\hbar \omega_c\approx 1$meV one needs $E/\hbar\lesssim 10^8$Hz. $E/\hbar$ has frequency units, corresponding to $10^8$ electrons/sec flowing into the box.

  While it may be possible to measure dissipation directly, e.g. via heating,
  we propose measuring instead the charge fluctuations (noise) $S_{Q}(\omega)=e^2\langle {\hat N}_t{\hat N}_{t'}\rangle_{\omega}$. The latter should be measured at
 frequency, temperature and level spacings $\Delta$ such that $\Delta<\omega, T \ll 10^8$Hz, to yield the  response to the force $E$. FDT relates the (symmetrized) noise and the retarded response $K(\omega)$ (Eq. \ref{e41}) via $S_Q(\omega)=\hbar\coth(\hbar\omega/2T)\im K(\omega)$. From Eq. (\ref{e45}) we have (at $T=0$) that the gate voltage averaged noise $\bar S_Q(\omega)$ satisfies $\bar S_{Q}(\omega)(\frac{2E_c}{e\hbar})^2\frac{1}{\omega}=\frac{\hbar}{\eta^R}$. In particular, as the fixed point is approached we predict $\bar S_{Q}(\omega)(\frac{2E_c}{e\hbar})^2\frac{1}{\omega}=2\pi$.

\vspace{1cm}

 Acknowledgements: We thank M. B\"{u}ttiker,  G. F\`{e}ve, Y. Gefen, A. Golub, D. Goldhaber-Gordon, K. Le Hur, S. L. Lukyanov, Y. Meir, C. Mora, B. Pla\c{c}ais and G. Zar\'{a}nd for stimulating discussions. BH acknowledges kind hospitality and financial support from LPTENS and PLD from Ben Gurion University. This research was supported by THE ISRAEL SCIENCE FOUNDATION (grant No. 1078/07) and by the ANR grant 09-BLAN-0097-01/2.

\appendix

\section{Mapping the Coulomb box and the ring }
The AES mapping has been extensively used, yet we find it useful to reproduce it since the relation between correlation functions has received less attention.

The Coulomb box action corresponding to the Hamiltonian (\ref{e40}) is
\beq{201}
-i\hbar S&=&\int_t\left\{\sum_{\alpha}d^{\dagger}_{\alpha, i}(i\hbar\partial_t-\epsilon_{\alpha})d_{\alpha, i} -E_c({\hat N}-N_0)^2\right\} -i\hbar S_{lead}-i\hbar S_{tun}\nonumber\\
-i\hbar S_{lead}&=&\int_t\sum_ka^{\dagger}_{k, i}(i\hbar\partial_t-\epsilon_k)a_{k,i}\nonumber\\
-i\hbar S_{tun}&=&\int_t\sum_{k,\alpha}t_{k,\alpha,i}a^{\dagger}_{k,i}d_{\alpha,i}+h.c.
\eeq
with the partition  $Z=\eexp{-S}$.
Adding a variable ${\dot {\theta}}_t$ to the path integral yields
\beq{202}
-i\hbar S&=&\int_t\left\{E_c[{\hat N}-N_0 -\frac{\hbar}{2E_c}{\dot {\theta}}_t]^2+
\sum_{\alpha}d^{\dagger}_{\alpha,i}(i\hbar\partial_t-\epsilon_{\alpha})d_{\alpha,i}
-E_c({\hat N}-N_0)^2\right\} -i\hbar S_{lead}-i\hbar S_{tun}\nonumber\\
&=&\int_t\left\{\sum_{\alpha}d^{\dagger}_{\alpha,i}(i\hbar\partial_t-\epsilon_{\alpha}-\hbar{\dot { \theta}}_t)d_{\alpha,i}+\frac{1}{4E_c}[\hbar\dot {\theta}_t+2E_cN_0]^2\right\}-i\hbar S_{lead}-i\hbar S_{tun}
\eeq
Now define  $d_{\alpha}=\eexp{-i\theta_t}
{\tilde d}_{\alpha}$
\beq{203}
-i\hbar S=\int_t\left\{\sum_{\alpha}{\tilde d}^{\dagger}_{\alpha,i}(i\hbar\partial_t-\epsilon_{\alpha}){\tilde d}_{\alpha,i}
+\frac{\hbar^2}{4E_c}{\dot\theta}_t^2+{\dot\theta}_tN_0+
\sum_{k,\alpha,i}[t_{k,\alpha,i}a^{\dagger}_{k,i}{\tilde d}_{\alpha,i}\eexp{i\theta_t}+h.c.]\right\}-i\hbar S_{lead}\nonumber\\
\eeq
The ring action in terms of $\theta_t$ is derived by integrating out the fermions ${\tilde d}_{\alpha}$ and $a_k$. Define time ordered Greens' functions on the dot $G_{0\alpha,i}(\omega)=\frac{1}{\omega-\epsilon_{\alpha,i}+i\text{sign}\omega 0^+}$ and on the lead
$G_{0k,i}(\omega)=\frac{1}{\omega-\epsilon_{k,i}+i\text{sign}\omega0^+}$. In matrix notation
\beq{204}
{\hat G}_i^{-1}(t,t')=\left(\begin{array}{cc}
G_{0\alpha,i}^{-1}(t,t') & 0\\
0 & G_{0k,i}^{-1}(t,t')\\
\end{array}\right)+\left(\begin{array}{cc}
0 & t_{k,\alpha,i}\eexp{i\theta_t}\\
t_{k,\alpha,i}^*\eexp{-i\theta_t} & 0\\
\end{array}\right)\delta(t-t')\equiv {\hat G}_{0i}^{-1}+{\hat T}_i
\eeq

The trace over fermions, using $\det(iG)=\eexp{Tr\ln iG}$, yields
\beq{205}
S_{eff}=-\sum_iTr\ln i{\hat G}_i^{-1}(t,t)=-\sum_iTr\ln \left\{i{\hat G}^{-1}_{0i}(t,t')[\delta(t-t')+
{\hat G}_{0i}(t',t){\hat T}_i(t)]\right\}
\eeq
Expanding in ${\hat T}$, the 0th order is $\theta_t$ independent, the 1st order vanishes, hence to 2nd order
\beq{206}
S_{eff}=-\half \sum_i Tr \{{\hat G}_0{\hat T}{\hat G}_0{\hat T}\}=
-\half\sum_i\int_{t,t'}G_{0\alpha,i}(t,t')G_{0k,i}(t',t)|t_{k,\alpha,i}|^2\eexp{i\theta_t-i\theta_{t'}}+h.c.
\eeq

For completeness we derive the Matsubara effective action using $\sum_{\alpha}G_{\alpha,i}(\tau)=T\sum_n G(\omega_n)\eexp{i\omega_n\tau}$ with fermionic $\omega_n=\pi T(2n+1)$,
\beq{207}
&&G(\omega_n)=\int_{\epsilon}\frac{\rho_{dot}(\epsilon)}{i\omega_n-\epsilon}=\int_0^{\infty}\rho_{dot}(\epsilon)
[\frac{1}{i\omega_n-\epsilon}+\frac{1}{i\omega_n+\epsilon}]=
\int_0^{\infty}\rho_{dot}(\epsilon)\frac{-2i\omega_n}{\omega_n^2+\epsilon^2}=-i\pi \rho_{dot}(0){\text sgn}(\omega_n)\nonumber\\&&
\sum_{\alpha}G_{0\alpha,i}(\tau)=2\pi \rho_{dot}(0)\sum_{n>0}\sin (\omega_n \tau)=\rho_{dot}(0)\frac{\pi T}{\sin (\pi T\tau)}
\eeq
where $\rho_{dot}(\epsilon)$ is the dot density of states, assumed symmetric, and eventually constant. With the lead density of states $\rho_{lead}(\epsilon)$, and assuming a constant $t_{k,\alpha,i}$
\beq{208}
S_{eff}=-\half|t|^2N_c\rho_{dot}(0)\rho_{lead}(0)\int\int\frac{\pi^2 T^2}{\sin^2[\pi T(\tau-\tau')]}
\cos[\theta(\tau)-\theta(\tau')]
\eeq
where $N_c=\sum_i$ is the number of channels. This is the well known equilibrium ring system with a bosonic CL environment \cite{guinea,golubev1,kagalovsky,bh} where
$\eta=\half\pi|t|^2N_c\rho_{dot}(0)\rho_{lead}(0)$ amd $m=1/2E_c$. The expansion in $\hat T$ is justified for $|t|^2\rightarrow 0$, however with $N_c\rightarrow\infty$ any value of $\eta$ can be generated. In fact $N_c$ can be fairly small and yet reproduce the $N_c\rightarrow\infty$ case, except at exponentially small temperatures \cite{zarand}.
A similar derivation holds for the Keldysh action leading to the form (\ref{e21}).

We proceed now to map observables of the Coloumb box to those of the ring problem. Since the action (\ref{e203}) has a term $+{\dot \theta} N_0$ we identify $N_0=-\phi_x$ where $\phi_x$ is the flux through the ring (in units of the quantum flux). Hence
\beq{209}
\hbar\langle {\dot \theta}\rangle&=&\int_{\theta} \hbar{\dot { \theta}}\eexp{-\frac{i}{\hbar}\int
E_c({\hat N}-N_0-\frac{\hbar}{2E_c}{\dot {\theta}})^2+\text{fermion terms}}\nonumber\\
&=&\int_{\theta} (\hbar{\dot {\theta}}+2E_c{\hat N}-2E_cN_0)\eexp{-\frac{i}{\hbar}\int
\frac{\hbar^2}{4E_c}{\dot {\theta}}^2+\text{fermion terms}}=2E_c[\langle{\hat N}\rangle-N_0]
\eeq
In particular, without interaction, $t_{k\alpha}=0$, the charge has no fluctuations $\langle{\hat N}\rangle=0$ (for $|N_0|<\half$) so that $\hbar\langle {\dot \theta}\rangle=-2E_cN_0=2E_c\phi_x$.

Consider next the time ordered ${\cal T}$ correlations (the following is the same for $\langle{\dot\theta}^+_t{\dot\theta}^+_{t'}\rangle, \langle{\dot\theta}^+_t{\dot\theta}^-_{t'}\rangle$ with $\pm$ Keldysh contours),
\beq{210}
\hbar^2{\cal T}\langle{\dot\theta}_t{\dot\theta}_{t'}\rangle&=&\int_{\tilde\theta} \hbar^2{\dot { \theta}}_t{\dot {\theta}}_{t'}\eexp{-\frac{i}{\hbar}\int
E_c({\hat N}-N_0-\frac{\hbar^2}{2E_c}{\dot {\theta}})^2+\text{fermion terms}}\nonumber\\
&=&\int_{\theta} (\hbar{\dot {\theta}}_t+2E_c{\hat N}_t-2E_cN_0)(\hbar{\dot {\theta}}_{t'}+2E_c{\hat N}_{t'}-2E_cN_0)\eexp{-\frac{i}{\hbar}\int
\frac{\hbar^2}{4E_c}{\dot {\tilde \theta}}^2+\text{fermion terms}}\nonumber\\
&=& \hbar^2{\cal T}\langle{\dot\theta}_t{\dot\theta}_{t'}\rangle_0+4E_c^2{\cal T}
\langle ({\hat N}_t-N_0)({\hat N}_{t'}-N_0)\rangle
\eeq
To obtain the retarded response,
\beq{211}
-i{\cal D}^R_{t,t'}=\theta(t-t')\langle [A_t,B_{t'}]\rangle=\theta(t-t')\langle A_tB_{t'}-B_{t'}A_t\rangle
={\text T}\langle A_t^+B_{t'}^+\rangle -\langle B_{t'}^-A_t^+\rangle
\eeq
where $\pm$ are Keldysh contour indices, so that $A^+$ is earlier than $B^-$.

 Define the response $K_{t,t'}$ of the Coulomb box, as well as the response of ring problem
${\tilde K}_{t,t'}$ in the form (displayed here with operators whose $\langle A_t\rangle=0$ to allow relation with time ordering),
\beq{212}
{\tilde K}_{t,t'}&=&+i\theta(t-t')\langle [({\dot \theta}_t-\langle{\dot\theta}\rangle),({\dot\theta}_{t'}-\langle{\dot\theta}\rangle)]\rangle
  \nonumber\\  K_{t,t'}&=&+i\theta(t-t')\langle [({\hat N}_t-\langle{\hat N}\rangle),({\hat N}_{t'}-\langle{\hat N}\rangle)]\rangle
\eeq
Fron Eq. (\ref{e210}) we have
\beq{214}
&&\hbar^2{\cal T}\langle ({\dot \theta}_t-\langle{\dot\theta}\rangle)({\dot\theta}_{t'}-\langle{\dot\theta}\rangle)\rangle+
\hbar^2\langle{\dot\theta}\rangle^2=\nonumber\\ &&\hbar^2{\cal T}\langle{\dot\theta}_t{\dot\theta}_{t'}\rangle_0
+4E_c^2{\cal T}\langle ({\hat N}_t-\langle{\hat N}\rangle)({\hat N}_{t'}-\langle{\hat N}\rangle)\rangle
+4E_c^2(\langle {\hat N}\rangle^2-2N_0\langle{\hat N}\rangle+N_0^2)
\eeq
Now using (\ref{e209}) and that the relation (\ref{e210}) holds for both terms in (\ref{e211}), a relation between these response functions is obtained
\beq{215}
\hbar^2{\tilde K}_{t,t'}&=&-2E_c\hbar\delta(t-t')+4E_c^2K_{t,t'}
\eeq
which is reproduced as Eq. (\ref{e40}). This relation is consistent with results in Ref. \onlinecite{burmistrov2}.

\section{Semiclassical case: 1st and 2nd order}

\subsection{ 1st order term}

First order perturbation of the Green's function
\be{0}
&\ & R^{(1)}_{t,t'} =-i\frac12 \int_{t1,t2} B_{t1,t2}
\avg{\hat\theta_{t1}\hat\theta_{t2} \cos(\theta_{t1}-\theta_{t2})\hat{\theta}_{t'}\theta_t}_{S_0}
= \\ \nonumber
&\ & \frac{-i}4 \int_{t1,t2} B_{t1,t2}
\sum_{\sigma=\pm} \partial_{\alpha_{i=1,2,3,4}} \  \Exp{i\alpha_1\hat\theta_{t1} +
i\alpha_2 \hat\theta_{t2} + i \sigma \theta_{t1}- i \sigma \theta_{t2} +
i\alpha_3 \hat\theta_{t'} + i \alpha_4 \theta_{t}}
\vert_{\alpha_i=0}
\ee

%
An Averaging with Gaussian weight
\be{0}
&& \avg{\eexp{i\theta_{t1}+i\theta_{t2}+...+i\hat\theta_{t1}+i\hat\theta_{t2}+...}} =
\eexp{i\avg{\theta_{t1}+\theta_{t2}+...}} \
\eexp{-\avg{(\theta_{t1}+\theta_{t2}+...)(\hat\theta_{t1}+\hat\theta_{t2}+...)}}
= \nonumber \\
&& \eexp{iv {t_1}+iv{t_2}+...} \eexp{i R_{t_1,t_2}+i R_{t_2,t_1} + ...}.
\ee
The retarded function
\be{0}
&& R^{(1)}_{t,t'} =  \nonumber\\
&& \frac1{4i} \int_{t_1,t_2} \sum_{\sigma=\pm} \partial_{\alpha_i}
B_{t_1,t_2} \ \eexp{i\alpha_1(-\sigma R_{t_2,t_1}+\alpha_4 R_{t,t_1})+
i\alpha_2(\sigma R_{t_1,t_2}-\alpha_4 R_{t,t_1})+
i\alpha_3(\sigma R_{t_1,t'}-\sigma R_{t_2,t'}+\alpha_4 R_{t,t_1})}\eexp{i\sigma v(t_1-t_2)} = \nonumber\\
&& \frac14 \int_{t_1,t_2} \sum_{\sigma=\pm} \partial_{\alpha_4}
B_{t_1,t_2} {(\sigma R_{t_2,t_1}-\alpha_4 R_{t,t_1})
(\sigma R_{t_1,t_2}+\alpha_4 R_{t,t_1})}
(\sigma R_{t_1,t'}-\sigma R_{t_2,t'}+\alpha_4 R_{t,t_1})
\eexp{i\sigma v(t_1-t_2)} = \nonumber \\
&& -\int_{t_1,t_2} B_{t_1,t_2} \cos v(t_1-t_2) R_{t,t_1} R_{t_1,t_2}
(R_{t_1,t'}-R_{t_2,t'})
\ee
In the last expression we use $R_t R_{-t}= 0$.

\subsection{2nd order term}

Using the same procedure for the second order
\be{0}
&\ & R^{(2)}_{t,t'} =
\frac i2 \avg{\hat\theta_{t'}\theta_t (S_{int})^2} = \\ \nonumber
&\ & -\frac i8 \int_{t_1,t_2,t_3,t_4} B_{t_1,t_2} B_{t_3,t_4}
\avg{\hat\theta_{t_1}\hat\theta_{t_2} \cos(\theta_{t_1}-\theta_{t_2})
\hat\theta_{t_3}\hat\theta_{t_4} \cos(\theta_{t_3}-\theta_{t_4}) \hat{\theta}_{t'}\theta_t} =  \frac1{2^5 i} \int_{t_{1..4}} B_{t_1,t_2} B_{t_3,t_4}  \nonumber \\
&\ &
\times \sum_{\sigma_1,\sigma_2=\pm} \partial_{\alpha_{i=1..6}} \avg{ \eexp{i\alpha_1\hat\theta_{t_1} + i\alpha_2 \hat\theta_{t_2} +
i\alpha_3\hat\theta_{t_3} + i\alpha_4 \hat\theta_{t_4} +
i \sigma_1 \theta_{t_1}- i \sigma_1 \theta_{t_2} +
i \sigma_2 \theta_{t_3}- i \sigma_2 \theta_{t_4} +
i\alpha_3 \hat\theta_{t'} + i \alpha_4 \theta_{t}} }
\vert_{\alpha_i=0}  \nonumber
\ee
using the symmetry between $\sigma_1 \leftrightarrow -\sigma_1$ and $t_1 \leftrightarrow t_2$ and similarly for $t_3,t_4$
\be{0}
&\ & R^{(2)}_{t,t'} =
\frac18 \int_{t_1,t_2,t_3,t_4} B_{t_1,t_2} B_{t_3,t_4} \eexp{iv(t_1-t_2)-iv(t_3-t_4)}
\partial_{\alpha_6} \left[-R_{t_2,t_1}+R_{t_3,t_1}-R_{t_4,t_1}
+\alpha_6 R_{t,t_1}\right] \nonumber\\
&& \left[R_{t_1,t_2}+R_{t_3,t_2}-R_{t_4,t_2}+\alpha_6 R_{t,t_2}\right]
\left[R_{t_1,t_3}-R_{t_2,t_3}-R_{t_4,t_3}+\alpha_6 R_{t,t_3}\right]\nonumber \\
&& \left[R_{t_1,t_4}-R_{t_2,t_4}+R_{t_3,t_4}+\alpha_6 R_{t,t_4}\right]
\left[R_{t_1,t'}-R_{t_2,t'}+R_{t_3,t'}-R_{t_4,t'}+\alpha_6 R_{t,t'}\right]	
\ee
the choice $t_1>t_2,t_3,t_4$, only $R_{t,t_1}$ remains. $R_\tau$ is real, we separate the exponent to two sinus and two cosine terms as follow
\be{0}
\label{R_2(t)}
&& R^{(2)}_{t,t'} = 
 \frac18 \int_{t_1,t_2,t_3,t_4} B_{t_1,t_2} B_{t_3,t_4}
\left(\cos v(t_1-t_2)\cos v(t_3-t_4)-\sin v(t_1-t_2)\sin v(t_3-t_4)\right)
R_{t,t_1} \nonumber \\
&& \left[R_{t_1,t_2}+R_{t_3,t_2}-R_{t_4,t_2}\right]
\left[R_{t_1,t_3}-R_{t_2,t_3}-R_{t_4,t_3}\right]
\left[R_{t_1,t_4}-R_{t_2,t_4}+R_{t_3,t_4}\right]  \nonumber\\
&& \left[R_{t_1,t'}-R_{t_2,t'}+R_{t_3,t'}-R_{t_4,t'}\right]	
\ee
This long multiplicity of $R_t$ terms is now separated to 8 different terms.
For the terms with the cosine we calculate explicitly
3 terms, which we label by $a$ to $c$. Term 'a':
\be{0}
&\ & R^{a}_{t,t'} = \frac12
\int_{t_1,t_2,t_3,t_4} B_{t_1,t_2} \cos v(t_1-t_2) \times\nonumber\\
&\ &  R_{t,t_1} R_{t_1,t_2}
(R_{t_1,t'}-R_{t_2,t'}) B_{t_3,t_4} \cos v(t_3-t_4) (R_{t_1,t_3}-R_{t_2,t_3})(R_{t_1,t_4}-R_{t_2,t_4}) =\nonumber \\
&\ & \frac12 \int_{t_1,t_2} B_{t_1,t_2} \cos v(t_1-t_2) R_{t,t_1} R_{t_1,t_2} (R_{t_1,t'}-R_{t_2,t'})
\ \tilde{C}_{t_1,t_2}
\ee
This term in $\omega$ space
\be{0}
&\ & R^{a}_\omega = -\frac12 R_\omega^2 \int_t R_t B_t \cos vt \
(\eexp{i\omega t}-1) \ \tilde{C}_t \nonumber
\ee
with $\tilde{C}_t=2(C^{(1)}_{t=0}-C^{(1)}_t)$.
Similarly we choose two different terms 'b' and 'c' and write them directly in $\omega$ space
\be{0}
&\ & R^{b}_\omega = R_\omega^2 \int_t R^{(1)}_t B_t \cos vt \
(\eexp{i\omega t}-1) \\
&\ & R^{c}_\omega = R_\omega^3 \left[\int_t R_t B_t \cos vt \
(\eexp{i\omega t}-1) \right]^2 = R_\omega^{-1} (R^{(1)}_\omega)^2 \
\ee
note the $R^{(1)}_t$ in the expression $R^{b}$ is the first order result of the retarded green function. $R^{c}_\omega$ is the reducible term containing multiplication of $R^{(1)}_\omega$. Renormalized $\eta$ for small $v$ is
\be{0}
&\ & \frac1{\eta_2^a} = \frac12 \frac1{\eta^2} \int_t R_t B_t \tilde{C}(t) \ t =
\frac{\hbar}{\pi\eta^3} \int_t R_t B_t \ t \left(\ln t +\gamma + \order{v} +\order{1/t} \right) = 
- \frac{\hbar^2}{2\pi^2\eta^3} \ln^2 v + \order{v}  \nonumber \\
&\ & \frac1{\eta_2^b} =  -\frac{\hbar}{\pi\eta^2} \int_t R^{(1)}_t B_t \ t =
-\frac{\hbar}{\pi\eta^3} \int_t R_t B_t \ t \left(\ln t +\gamma + 1 + \order{v} +\order{1/t} \right) = \nonumber \\
&\ & \frac{\hbar^2}{2\pi^2\eta^3} \ln^2 v - \frac{\hbar^2}{2\pi^2\eta^3} \ln v + \order{v} \nonumber  \\
&\ & \frac1{\eta_2^c} =  \frac1{\eta^3} \left[\int_t R_t B_t \ t \right]^2  =
\frac{\hbar^2}{2\pi^2\eta^3} \left[ \ln v + \order{v} \right]^2 =
\frac{\hbar^2}{2\pi^2\eta^3} \ln^2 v + \order{v}
\ee

The terms containing the sine in Eq. (\ref{R_2(t)}), are in general of order $\order{v}$, however we have identify the following term which, depending on the order of limits,
may contribute a term logarithmic in $v$ for small $v$.
\be{0}
 R^{d}_\omega &=& - R_\omega^2 \int_{t_1,t_2} R_{t_1} R_{t_2} B_{t_1} B_{t_2}
\sin v t_1 \sin v t_2
(1-\eexp{i \omega t_1}) \int_{t_3} \left(R_{t_1+t_3}-R_{t_3}\right)
\ee

We label the dissipation parameter form this term by $\delta(\frac1{\eta_2^R}) = 
\lim_{\omega\rightarrow0} (-i\omega)  R^{d}_\omega$ and found the logarithmic prefactor in Eq. (\ref{eta_d}), where we use for $t_1>0$
\be{0}
\int_{t_3} \left(R_{t_1+t_3}-R_{t_3}\right) =
\frac1\eta \int_{-t_1}^0 \left(1-\eexp{-(t_1+t_3) \frac\eta m} \right) +
\frac1\eta \int_0^\infty \left(\eexp{-t_3 \frac\eta m} - \eexp{-(t_1+t_3) \frac\eta m} \right)= \frac{t_1}\eta
\ee

%
%
%
%
%

\section{Quantum case: 1st order, more details}
\label{app:integral}

Let us give the detailed calculation of the first order correction in the case of a
mass only cutoff, i.e. $\tau_0=0$. Taking the derivative of Eq. (\ref{eq63}) in the text
we have:
\begin{eqnarray}
 \partial_v \delta E^{(1)} &=&  - \frac{2}{\hbar} \int_{\tau > 0} \tau B(\tau)  \sin( \frac{\hbar}{2} R(\tau)) \cos(v \tau)  =   \frac{2 \eta}{\pi}  \int_{\tau > 0}  \frac{d \tau}{\tau} \sin( \frac{\hbar}{2 \eta} (1-e^{-\frac{\eta}{m} \tau}) ) \cos(v \tau) \nonumber\\
&=&  \frac{2 \eta}{\pi} [ \sin( \frac{\hbar}{2 \eta}) \int_{\tau > 0}  \frac{d \tau}{\tau}  (1-e^{-\frac{\eta}{m} \tau})  \cos(v \tau)
\nonumber\\ && -   \int_{\tau > 0}  \frac{d \tau}{\tau} [ \sin( \frac{\hbar}{2 \eta} (1-e^{-\frac{\eta}{m} \tau}) ) -  \sin( \frac{\hbar}{2 \eta}) (1-e^{-\frac{\eta}{m} \tau}) ] \cos(v \tau) ] \nonumber \\
&=&   \frac{2 \eta}{\pi} [ \sin( \frac{\hbar}{2 \eta})   \ln(\frac{\eta}{m v}) +
f(\frac{\hbar}{2 \eta}) + O(1/v) ]
\end{eqnarray}
since the first integral can be computed exactly and in the second one can set $v=0$ to get the constant piece.
This determines the constant $C= f(\frac{\hbar}{2 \eta})  $ given in the text in Eq. (\ref{eq64}), where the
function $f(x)$ is defined as:
\begin{eqnarray}
&& f(x) =  \int_{0}^{+\infty}  \frac{d t}{t} [ \sin(x(1-e^{- t}) ) -  \sin(x) (1-e^{- t}) ] \\
&& = - \int_0^1 \frac{dz}{(1-z) \ln(1-z)} (\sin (x z) - z \sin x)  = \frac{1}{6} x^3 \ln \left(\frac{8}{3}\right)+ O(x^5)  \nonumber
\end{eqnarray}
and is a nicely convergent integral, where one can rescale $t$ freely. Although it is not periodic in $x$, upon plotting it one notes that it seems to
become almost periodic a large $x$.

\section{Quantum case: 2nd order for $\tau_1\rightarrow 0$}

Since $\sin(\half\hbar R_{t_1,t_2})$ is a retarded function, we use
for $R_t = \Theta(t)\eexp{-\delta t}$
\beq{301}
\sin(\half\hbar R_{t_1,t_2})&&\rightarrow \sin(\frac{\hbar}{2\eta})\eexp{-\delta(t_1-t_2)}\nonumber\\
\sin[\half\hbar(R_{t_1,t_3}+R_{t_2,t_3})]&&\rightarrow \sin(\frac{\hbar}{\eta})\eexp{-\delta(t_1-t_3)-\delta(t_2-t_3)}\nonumber\\
\sin[\half\hbar(R_{t_1,t_4}-R_{t_3,t_4}+R_{t_2,t_4})]&&\rightarrow \sin(\frac{\hbar}{2\eta})\eexp{-\delta(t_1-t_4)-\delta(t_3-t_4)-\delta(t_2-t_4)}
\eeq
e.g.  Fourier of $t_1-t_3$ and $t_2-t_3$ should have  $1/(\omega_1+i\delta)(\omega_2+i\delta)$. define the variables
\beq{302}
t_2'&=&t_2-t_1,\qquad t_3'=t_3-t_2,\qquad t_4'=t_4-t_3\nonumber\\
\Rightarrow \qquad t_2&=&t_2'+t_1,\qquad t_3=t_3'+t_2'+t_1, \qquad t_4=t_4'+t_3'+t_2'+t_1
\eeq
These variables are more convenient since their range is independent $-\infty<t_2',t_3',t_4'<0$. The product of all convergence factors is then $\eexp{\delta(3t_2'+4t_3'+3t_4')}$, with 3,4,3 factors unimportant since $\delta\rightarrow 0$.
Hence
\beq{303}
&&\delta E^{(2)}=\frac{4}{\hbar^3}\sin^2(\frac{\hbar}{2\eta})\sin(\frac{\hbar}{\eta})
\int_{\omega_1,\omega_2}B(\omega_1)B(\omega_2)\sum_{\sigma=\pm}\frac{\sigma}{2i}\int_A
\eexp{i\sigma v(-2t_3'-t_4'-t_2')}\\
&&\times [\eexp{i\omega_1(t_3'+t_2')+i\omega_2(t_4'+t_3')}
+\eexp{i\omega_1(t_4'+t_3'+t_2')+i\omega_2t_3'}]\eexp{\delta(t_2'+t_3'+t_4')} = \nonumber\\
&&\frac{4}{\hbar^3}\sin^2(\frac{\hbar}{2\eta})\sin(\frac{\hbar}{\eta})
\int_{\omega_1,\omega_2}B(\omega_1)B(\omega_2)\sum_{\sigma=\pm}\frac{\sigma}{2i}\times\nonumber\\
&&[ \frac{1}{-i\sigma v+i\omega_2+\delta}+\frac{1}{-i\sigma v+i\omega_1+\delta}]
\frac{1}{(-2i\sigma v+i\omega_1+i\omega_2+\delta)(-i\sigma v+i\omega_1+\delta)} = \nonumber\\
&&\frac{4}{\hbar^3}\sin^2(\frac{\hbar}{2\eta})\sin(\frac{\hbar}{\eta})
\sum_{\sigma}\frac{\sigma}{2}(\hbar\eta)^2\int\frac{d\omega_1}{2\pi}\frac{1}{(\omega_1-\sigma v-i\delta)^2}\frac{|\omega_1|}{1+\omega_1^2\tau_0^2}\int\frac{d\omega_2}{2\pi}
\frac{1}{\omega_2-\sigma v-i\delta}\frac{|\omega_2|}{1+\omega_2^2\tau_0^2}\nonumber
\eeq
with the integral over $\omega_2$
\beq{304}
&&\int_0^{\infty}d\omega_2 [\frac{1}{\omega_2-\sigma v-i\delta}-\frac{1}{-\omega_2-\sigma v-i\delta}]\frac{\omega_2}{1+\omega_2^2\tau_0^2}=2\sigma v\int_0^{\infty} d\omega_2
\frac{\omega_2}{(\omega_2^2-v^2)(1+\omega_2^2\tau_0^2)}\nonumber\\
&&=-\sigma v \ln(v\tau_0) +O(v^3\tau_0^2\ln(v\tau_0)
\eeq
and over $\omega_1$
\beq{305}
&&\int_0^{\infty}d\omega_1 [\frac{1}{(\omega_1-\sigma v-i\delta)^2}+\frac{1}{(-\omega_1-\sigma v-i\delta)^2}]\frac{\omega_1}{1+\omega_1^2\tau_0^2}\nonumber\\
&&=2\int_0^{\infty}d\omega_1
[\frac{\omega_1}{\omega_1^2-v^2}+\frac{2v^2\omega_1}{(\omega_1-\sigma v-i\delta)^2(\omega_1+\sigma v+i\delta)^2}]=-2\ln(v\tau_0)-2
\eeq
where in the last integral $\tau_0\rightarrow 0$ can be taken. Substituting (\ref{e304},\ref{e305}) in (\ref{e303}) leads  to the result Eq. (\ref{e70}).

\section{Quantum case: 2nd order with a mass cutoff}

In this appendix we rederive the 2nd order quantum case using directly a mass cutoff. In particular we identify the coefficient of  the
$\ln^2v$ term, confirming that coefficient in Eq. (\ref{e70}), and derive some properties of the 2nd order $\ln v$ term.

We express Eq. \ref{e66} as

\beq{401}
\delta E^{(2)} = \frac{i}{4\hbar^3}\sum_{\epsilon_i}\epsilon_2\epsilon_3\epsilon_4
\int_{t_2,t_3,t_4}B_{t_1,t_2}B_{t_3,t_4}A_2\sin[v(t_1-t_2+ v(t_3-t_4)]
\eeq
where the symmetry between $t_3$ and $t_4$ is used
to sum over $\mu=\pm$. Defining
$ F_{t_i,t_j} = \eexp{i\epsilon \frac\hbar2 R_{t_i,t_j}} -1 $
Eq. \ref{e67} can be expressed as
\be{0}
A_2=
&&(F^{-\epsilon_1}_{t_2,t_1}+1)(F^{\epsilon_1}_{t_3,t_1}+1)(F^{-\epsilon_1}_{t_4,t_1}+1)
(F^{\epsilon_2}_{t_1,t_2}+1)(F^{\epsilon_2}_{t_3,t_2}+1)(F^{-\epsilon_2}_{t_4,t_2}+1)\nonumber \\
&&(F^{\epsilon_3}_{t_1,t_3}+1)(F^{-\epsilon_3}_{t_2,t_3}+1)(F^{-\epsilon_3}_{t_4,t_3}+1)
(F^{\epsilon_4}_{t_1,t_4}+1)(F^{-\epsilon_4}_{t_2,t_4}+1)(F^{\epsilon_4}_{t_3,t_4}+1) = \nonumber\\
&& (F^{\epsilon_2}_{t_1,t_2}+F^{-\epsilon_1}_{t_2,t_1}+1)
(F^{\epsilon_3}_{t_1,t_3}+F^{\epsilon_1}_{t_3,t_1}+1)
(F^{\epsilon_4}_{t_1,t_4}+F^{-\epsilon_1}_{t_4,t_1}+1) \nonumber \\
&& (F^{\epsilon_2}_{t_3,t_2}+F^{-\epsilon_3}_{t_2,t_3}+1)
(F^{-\epsilon_2}_{t_4,t_2}+F^{-\epsilon_4}_{t_2,t_4}+1)
(F^{-\epsilon_3}_{t_4,t_3}+F^{\epsilon_4}_{t_3,t_4}+1)
\ee
In the last expression we used
the retarded property of $F_{t_i,t_j}$
so that $F_{t_i,t_j}F_{t_j,t_i} = 0$.
When transforming all functions to their frequency domain
\be{0}
\delta E^{(2)} &=& \frac{1}{16\hbar^3}
\int_{\omega_a,\omega_b}
\left( [B_{\omega_a+v} + B_{\omega_a-v} ][ B_{\omega_b+v} - B_{\omega_b-v}]  + \right. \nonumber \\
&& \left. [B_{\omega_a+v} - B_{\omega_a-v} ][ B_{\omega_b+v} + B_{\omega_b-v}]\right)
 K(\omega_a,\omega_b)
 \ee
\beq{404}
  K(\omega_a,\omega_b) &=&  \sum_{\epsilon_i}\epsilon_2\epsilon_3\epsilon_4\int_{\omega_1,..\omega_6}
  [F^{\epsilon_2}_{\omega_1} + F^{-\epsilon_1}_{-\omega_1} +2\pi\delta(\omega_1) ]
  [F^{\epsilon_3}_{\omega_2} + F^{\epsilon_1}_{-\omega_2} +2\pi\delta(\omega_2) ]
  [F^{\epsilon_4}_{\omega_3} + F^{-\epsilon_1}_{-\omega_3} +2\pi\delta(\omega_3) ]\nonumber \\
&&[F^{\epsilon_2}_{\omega_4} + F^{-\epsilon_3}_{-\omega_4} +2\pi\delta(\omega_4) ]
   [F^{-\epsilon_2}_{\omega_5} + F^{-\epsilon_4}_{-\omega_5} +2\pi\delta(\omega_5) ]
   [F^{-\epsilon_3}_{\omega_6} + F^{\epsilon_4}_{-\omega_6} +2\pi\delta(\omega_6) ] \nonumber \\
&& (2\pi)^3\delta(\omega_a + \omega_1 + \omega_4 + \omega_5)
   \delta(-\omega_b + \omega_2 - \omega_4 + \omega_6)
   \delta(\omega_b + \omega_3 - \omega_5 - \omega_6)
 \eeq
We notice that the function $K(\omega_a,\omega_b)$
can have poles at $\omega_a,\omega_b=i\delta$ leading to
a logarithmic divergence term for either a
$\order{\omega^{-1}}$ term with the antisymmetric expression
\beq{405}
\int_{\omega} ( [B_{\omega+v} - B_{\omega-v} ) \frac{1}{\omega-i\delta} =
-2 \int_0^{\infty}  B(\tau) \sin (v\tau) d \tau = \frac{2\hbar\eta}\pi v \ln(v) + \order{v}
\eeq
or for $\order{\omega^{-2}}$ terms with the symmetric expression
\beq{406}
\int_{\omega} ( [B_{\omega+v} + B_{\omega-v}) \frac{1}{(\omega-i\delta)^2} =
-2 \int_0^{\infty} \tau \Theta(\tau) B(\tau) \cos(v\tau) d\tau
= \frac{2\hbar\eta}\pi \ln(v) +\order{v}
\eeq
where $\delta=+0$. Note that the Fourier transform of $1/(\omega-i\delta)$ is
$\eexp{-\delta \tau} \Theta{\tau} $ while that of $1/(\omega-i\delta)^2$ is
$\eexp{-\delta \tau} \tau \Theta{\tau} $. We keep here only the long time divergence, controlled by $\ln v$.
Keeping also short time divergences would eventually replace $\ln v\rightarrow \ln \frac{v}{\omega_c}$ with $\omega_c=\eta/m$.
Eqs. (\ref{e405},\ref{e406}) show that $\ln^2(v)$ terms arises from either a $1/{\omega_a\omega_b^2}$
or $1/{\omega_a^2\omega_b}$ terms in $K(\omega_a,\omega_b)$.

We use the retarded property of $F_\tau = F_\tau\Theta(\tau)$ and expand the
function in power of $\hbar/\eta$
\be{0}
F_{\omega}^{\epsilon} =
\eexp{i\epsilon \frac{\hbar}{2\eta}}
\sum_{n=0}^\infty \frac1{n!} \left(-\frac{i \hbar \epsilon}{2\eta}\right)^n
\frac i{\omega  + i n \omega_c + i\delta} - \frac i{\omega+i\delta}
\ee
Each of the six factors takes the form
\beq{407}
F^{\epsilon_i}_\omega + F^{\epsilon_j}_{-\omega} +2\pi \delta(\omega)=
\sum_{n=0}^\infty \frac1{n!} \left(-\frac{i \hbar }{2\eta}\right)^n
\left\{
\frac {i\epsilon_i^n \eexp{i\epsilon_i \frac{\hbar}{2\eta}} }{\omega  + i n \omega_c + i\delta} +
\frac {i\epsilon_j^n \eexp{i\epsilon_j \frac{\hbar}{2\eta}} }{-\omega  + i n \omega_c + i\delta} \right\}
\eeq
where the delta function cancel with the last terms of the $F_\omega$'s.
We note that $\ln v$ terms arise from terms with at least one vanishing $n_j$, leading to a pole. For that particular $n_j$ the pole has a coefficient $\eexp{i\epsilon_i \frac{\hbar}{2\eta}}-\eexp{i\epsilon_j \frac{\hbar}{2\eta}}$ that vanishes when $\frac{\hbar}{2\eta}=\pi\times$integer. Hence all terms of $\delta E^{(2)}$ have at least one periodic factor of $\sin \frac{\hbar}{2\eta}$.

The triple frequency integral Eq. (\ref{e404}) with the substitution (\ref{e407}) has 24 terms all with three poles in either $\omega_a$ or $\omega_b$.
Solving for the triple integral and the $\epsilon_j$ summations we find
\be{0}
&&  K(\omega_a,\omega_b) =  \sum_{n_1,..n_6\ge0}
\frac1{n_1!n_2!n_3!n_4!n_5!n_6!} \left(-\frac{i \hbar }{2\eta}\right)^{n_1+n_2+n_3+n_4+n_5+n_6}
 \times  \nonumber \frac2{\omega_c^3}   \\
&&
\left\{ \frac{  ( (-1)^{n_2} - (-1)^{n_4})
 ((-1)^{n_1} - \eexp{-i \frac\hbar\eta} )}
{ (n_2+n_3+n_4+n_5+\delta)(n_1+n_2+n_3+ \delta + i\omega_a/\omega_c) }
 \left( \frac{(-1)^{n_3} - (-1)^{n_5+n_6} \eexp{i \frac\hbar\eta} } {n_3+n_5+n_6+\delta-i\omega_b/\omega_c} +
 \right. \right. \nonumber \\
&& \left.
\frac{(-1)^{n_5} - (-1)^{n_3+n_6} \eexp{-i \frac\hbar\eta}} {n_3+n_5+n_6+\delta i\omega_b/\omega_c} \right)  +
\nonumber \\
&&  \frac{ (-1)^{n_2} - \eexp{-i \frac\hbar\eta} }
{(n_1+n_2+n_3+ \delta + i\omega_a/\omega_c)}
\left( \frac{ ( (-1)^{n_3} - (-1)^{n_5+n_6} \eexp{i \frac\hbar\eta} )
( (-1)^{n_1} - (-1)^{n_4})}
{(n_3+n_5+n_6+\delta-i\omega_b/\omega_c)
(n_1+n_3+n_4+n_6+\delta+i(\omega_a-\omega_b)/\omega_c)} +
\right.   \nonumber \\
&& \left. \frac{ ( (-1)^{n_3+n_6} \eexp{i \frac\hbar\eta}- (-1)^{n_5}  )
( (-1)^{n_1+n_4} \eexp{i \frac\hbar\eta} -
\eexp{-i \frac\hbar\eta} )}
{(n_3+n_5+n_6+\delta+i\omega_b/\omega_c)
(n_1+n_3+n_4+n_6+\delta+i(\omega_a+\omega_b)/\omega_c)} \right) + \nonumber \\
&&  \frac{ (-1)^{n_2} - \eexp{-i \frac\hbar\eta} }
{(n_1+n_2+n_3+ \delta + i\omega_a/\omega_c)
(n_1+n_4+n_5+ \delta + i\omega_a/\omega_c)}
\left( \frac{ ( (-1)^{n_3} - (-1)^{n_6}  )
( (-1)^{n_1+n_5} \eexp{i \frac\hbar\eta} - (-1)^{n_4})}
{(n_1+n_3+n_4+n_6+\delta+i(\omega_a-\omega_b)/\omega_c)} +
\right.   \nonumber \\
&& \left.\left. \frac{ ( (-1)^{n_1+n_4} \eexp{i \frac\hbar\eta} - (-1)^{n_5}  )
( (-1)^{n_3+n_6} \eexp{i \frac\hbar\eta} -
\eexp{-i \frac\hbar\eta} )}
{(n_1+n_3+n_4+n_6+\delta+i(\omega_a+\omega_b)/\omega_c)} \right) \right\}
\ee

At this stage the $\ln^2v$ term can be simply identified, since this term needs poles in both $\omega_a$ and $\omega_b$. The only such term which has the form $\frac{1}{(\omega_a-i\delta)(\omega_b-i\delta)^2}$, is the term where $n_1=n_2=...=n_6=0$; all
other terms do not have a zero frequency divergence at both $\omega_a$ and $\omega_b$.
For this term we get
\be{0}
K_0(\omega_a,\omega_b) = \frac{
16  \sin^2 \frac{\hbar}{2\eta} \sin \frac\hbar\eta}
{(\omega_a - i \delta\omega_c)^2 (\omega_b -i \delta\omega_c)}
\ee
And the full expression from Eq. (\ref{e401}), using Eqs. ({\ref{e405},\ref{e406}), is then
\be{0}
\delta E^{(2)} =&& \frac{16}{16\hbar^3} \times \frac{4\hbar^2\eta^2}{\pi^2}
v \ln^2(v)  \times \sin^2 \frac{\hbar}{2\eta} \sin \frac\hbar\eta +\order{\ln v}=
\nonumber \\
&& \frac{4 \eta^2}{\pi^2\hbar} \sin^2 \frac{\hbar}{2\eta} \sin \frac\hbar\eta\cdot v\ln^2(v)+\order{\ln v}
\ee
This coefficient of the $v\ln^2(v)$ term agrees with that in Eq (\ref{e70}).

We consider next some of the terms that contribute to the $\ln v$ coefficient.
From Eq. (\ref{e405})  we know that only terms with a single pole , i.e. either
$1/(\omega_a-i\delta)$  or  $1/(\omega_b-i\delta)$, contribute. We define an expansion
\be{0}
&\ & K(\omega_a,\omega_b) =K_0(\omega_a,\omega_b)+
\sum_{\bar n=1}^{\infty}
\left(-\frac{i \hbar }{2\eta}\right)^{\bar n}\frac{2}{\omega_c^2}k_{\bar n}(\omega_a,\omega_b)
\ee
where $\bar n=\sum_{j=1}^6 n_j$. Thus
there are 6 terms for $\bar n=1$, 21 terms for $\bar n=2$ and 56 terms for $\bar n=3$. Due to the $\omega_a,\omega_b$ symmetry we define
\be{0}
&\ & \kappa_{\bar n}(\omega)  = \lim_{\omega_a \rightarrow 0} \omega_a k_{\bar n}(\omega_a,\omega) +\lim_{\omega_b \rightarrow 0} \omega_b k_{\bar n}(\omega,\omega_b)
\ee
so that one integration gives a $\ln v$ while the other gives its coefficient in the form
\be{0}
\delta E^{(2)}=\frac{4 \eta^2}{\pi^2\hbar} \sin^2 \frac{\hbar}{2\eta} \sin \frac\hbar\eta \cdot v\ln^2(v)
+\frac{\eta}{2\omega_c^2\hbar^2\pi}\sum_{\bar n=1}^{\infty}
\left(-\frac{i \hbar }{2\eta}\right)^{\bar n}
\int_{\omega} B_{\omega}\kappa_{\bar n}(\omega)\cdot v \ln(v) +\order 1
\ee

For the first few terms we find
\be{0}
&\ & \kappa_{1}(\omega) = \frac{P_2 + P_2 \cos\frac\hbar\eta  }{P_4} \sin^2\frac\hbar{2 \eta} \nonumber\\
&\ & \kappa_{2}(\omega) = \frac{P_4 + P_4 \cos\frac\hbar\eta  }{P_6} \sin\frac\hbar{2\eta}\nonumber\\
&\ & \kappa_{3}(\omega) = \frac{P_8 + P_8 \cos\frac\hbar\eta  }{P_{10}} \sin^2\frac\hbar{2 \eta}
\ee
where $P_I$ is a polynomial of $\omega/\omega_c$  of degree $I$. The result is consistent with having at least one factor of $\sin\frac\hbar{2\eta}$, as shown above in general.


\begin{thebibliography}{99}
\bibitem{buttiker} M. B\"{u}ttiker, H. Thomas, H., and A. Pr\^{e}tre,  Phys. Lett. A {\bf 180}, 364 (1993).
\bibitem{gabelli} J. Gabelli, et al.  Science {\bf 313}, 499 (2006).
\bibitem{mora} C. Mora and K. Le Hur, Nature Physics {\bf 6}, 697 (2010).
\bibitem{filippone} M. Filippone and C. Mora, arXiv:1205.2213.
\bibitem{kleemans} N. A. J. M. Kleemans et al, Phys. Rev. Lett. {\bf 99}, 146808 (2007).
\bibitem{guinea} F. Guinea, Phys. Rev. B{\bf 65}, 205317 (2002).
\bibitem{golubev1} D. S. Golubev, C. P. Herrero and A. D. Zaikin, Europhys. Lett. {\bf 63}, 426 (2003).
\bibitem{kagalovsky} V. Kagalovsky and B. Horovitz, Phys. Rev. B{\bf 78}, 125322 (2008).
\bibitem{bh} B. Horovitz and P. Le Doussal, Phys. Rev. B {\bf 74}, 073104 (2006) and Phys. Rev. B 82, 155127 (2010).
\bibitem{arrachea} L. Arrachea, Phys. Rev. B{\bf 66}, 045315 (2002); L. Arrachea, Phys. Rev. B{\bf 70}, 155407 (2004); F. Foieri, L. Arrachea and M. J. S\'{a}nchez, Phys. Rev. Lett. {\bf 99}, 266601 (2007).
\bibitem{AES} V. Ambegaokar, U. Eckern, and G. Sch\"{o}n, Phys. Rev. Lett. {\bf 48}, 1745 (1982);
Phys. Rev. B{\bf 30}, 6419 (1984).
\bibitem{CL} A. O. Caldeira and A. J. Leggett, Physica A {\bf 121}, 587 (1983).
\bibitem{schon} For a review see  G. Sch\"{o}n and A. D. Zaikin, Phys. Rep. {\bf 198}, 237 (1990).
\bibitem{golub} A. A. Golub, O. V. Grimalsky and Ya. I. Kerner, Europhys. Lett. {\bf 4}, 617 (1987).
\bibitem{falci} G. Falci, G. Sch\"{o}n and G. T. Zimanyi, Phys. Rev. Lett. {\bf 74}, 3257 (1995).
\bibitem{hofstetter} W. Hofstetter and W. Zwerger, Phys. Rev. Lett. {\bf 78}, 3737 (1997).
\bibitem{herrero} C. P. Herrero, G. Sch\"{o}n and A. D. Zaikin, Phys.
Rev. B{\bf 59}, 5728 (1999).
\bibitem{bulgadayev} S. A. Bulgadayev, JETP Lett. {\bf 83}, 563 (2006).
\bibitem{lukyanov1} S. L. Lukyanov and A. B. Zamolodchikov, J. Stat. Mech. P05003 (2004).
\bibitem{lukyanov2} S. L. Lukyanov and P. Werner, J. Stat. Mech. P11002 (2006).
\bibitem{burmistrov1} A. M. M. Pruisken and I. S. Burmistrov, Phys. Rev. Lett. {\bf 95},
189701 (2005); Phys. Rev. B{\bf 81}, 085428 (2010).
\bibitem{burmistrov2} Ya. I. Rodionov, I. S. Burmistrov and A. S. Ioselevich, Phys. Rev. B{\bf 80}, 035332 (2009).
\bibitem{buttiker2} M. B\"{u}ttiker, Annals N.Y. Acad. {\bf 480}, 194 (1986).
\bibitem{devoret} P. Joyez, D. Esteve and M. H. Devoret, Phys. Rev. Lett. {\bf 80}, 1956 (1998).
\bibitem{etzioni} Y. Etzioni, B. Horovitz and P. Le Doussal, Phys. Rev. Lett. {\bf 106}, 166803 (2011).
\bibitem{thouless} D. J. Thouless, Phys. Rev. B{\bf 27}, 6083 (1983).
\bibitem{ford} G. W. Ford, J. T. Lewis and R. F. O'Connell,  Phys. Rev. A{\bf 37}, 4419 (1988).
\bibitem{golubev2} D. S. Golubev and A. D. Zaikin, Phys. Rev. Lett. {\bf 86}, 4887 (2001).
\bibitem{hakim} V. Hakim and V. Ambegaokar, Phys. Rev. A{\bf 32}, 423 (1985).
\bibitem{fisher} M. P. A. Fisher and W. Zwerger, Phys. Rev. B{\bf 32}, 6190 (1985).
\bibitem{spohn} H. Spohn and W. Zwerger, J. Stat. Phys. {\bf 94}, 1037 (1999).
\bibitem{aleiner} I. L. Aleiner and A. V. Andreev, Phys. Rev. Lett. {\bf 81}, 1286 (1998).
\bibitem{sharma} P. Sharma and C. Chamon, Phys. Rev. Lett. {\bf 87}, 096401 (2001).
\bibitem{shimshoni} E. Shimshoni and Y. Gefen, Annals of Physics, {\bf 210}, 16 (1991).
\bibitem{zarand} G. Zar\'{a}nd, G. T. Zim\'{a}nyi and F. Wilhelm, Phys. Rev. B{\bf 62}, 8137 (2000).
\end{thebibliography}
\end{document}